%% file: EC_performance__arXiv.tex
\documentclass{scrartcl}

\usepackage[utf8]{luainputenc}
\usepackage[USenglish]{babel}
\usepackage{csquotes}

\usepackage[a4paper,top=27mm,bottom=20mm,inner=25mm,outer=20mm]{geometry}

\usepackage[%
  backend=bibtex,bibencoding=ascii,
  style=numeric-comp,
  giveninits=true, uniquename=init, 
  natbib=true,
  url=true,
  doi=true,
  isbn=false,
  backref=false,
  maxnames=99,
  ]{biblatex}
\usepackage{amsmath}
\allowdisplaybreaks
\numberwithin{equation}{section}
\usepackage{amssymb}
\usepackage{commath}
\usepackage{mathtools}
\usepackage{bbm}
\usepackage{nicefrac}
\usepackage{subdepth}

\usepackage{siunitx}
\usepackage{algorithm}
\usepackage{algpseudocode}

\usepackage{jlcode}

\lstdefinestyle{code}{
  language=Julia,
  showstringspaces=false,
  keywordstyle=\color{blue},
  commentstyle=\color{gray},
  identifierstyle=\color[RGB]{0,102,0},
  columns=fullflexible,
  keepspaces=true
}
\lstnewenvironment{code}{\lstset{style=code}}{}
\newcommand{\codeinline}{\jlinl}

\usepackage{amsthm}
\usepackage{thmtools}
\usepackage{etoolbox}
\makeatletter
\patchcmd{\thmt@setheadstyle}
 {\bgroup\thmt@space}
 {\thmt@space}
 {}{}
\patchcmd{\thmt@setheadstyle}
 {\egroup\fi}
 {\fi}
 {}{}
\makeatother
\declaretheoremstyle[
  bodyfont=\normalfont\itshape,
  headformat=\NAME\ \NUMBER\NOTE,
]{myplain}
\declaretheoremstyle[
  headformat=\NAME\ \NUMBER\NOTE,
]{mydefinition}
\newcommand{\envqed}{{\lower-0.3ex\hbox{$\triangleleft$}}}
\declaretheorem[style=myplain,numberwithin=section]{theorem}

\declaretheorem[style=mydefinition,numberlike=theorem,qed=\envqed]{definition}

\declaretheorem[style=mydefinition,numberlike=theorem,qed=\envqed]{example}

\usepackage[plainpages=false,pdfpagelabels,hidelinks,unicode]{hyperref}

\usepackage{color}
\usepackage{graphicx}
\usepackage[small]{caption}
\usepackage{subcaption}

\begingroup\expandafter\expandafter\expandafter\endgroup
\expandafter\ifx\csname pdfsuppresswarningpagegroup\endcsname\relax
\else
\fi

\usepackage{booktabs}
\usepackage{rotating}
\usepackage{multirow}

\usepackage{enumitem}

\usepackage{ifluatex}
\ifluatex
  \usepackage[no-math]{fontspec}
\else
  \usepackage[T1]{fontenc}
\fi
\usepackage{newpxtext,newpxmath}

\input{macros}

\newcommand{\orcid}[1]{ORCID:~\href{https://orcid.org/#1}{#1}}
\usepackage{authblk}

\newenvironment{keywords}{\par\textbf{Key words.}}{\par}

\title{\vspace{-1cm}Efficient implementation of modern entropy stable and kinetic energy preserving discontinuous Galerkin methods for conservation laws}

\author[1]{Hendrik~Ranocha\thanks{\orcid{0000-0002-3456-2277}}}
\affil[1]{Applied Mathematics, University of Hamburg, Germany}

\author[2]{Michael~Schlottke-Lakemper\thanks{\orcid{0000-0002-3195-2536}}}
\affil[2]{High-Performance Computing Center Stuttgart, University of Stuttgart, Germany}

\author[3]{Jesse~Chan\thanks{\orcid{0000-0003-2077-3636}}}
\affil[3]{Computational and Applied Mathematics, Rice University, USA}

\author[4]{Andr{\'e}s~M.~Rueda-Ram\'{i}rez\thanks{\orcid{0000-0001-6557-9162}}}
\affil[4]{Department of Mathematics and Computer Science, University of Cologne, Germany}

\author[5]{Andrew~R.~Winters\thanks{\orcid{0000-0002-5902-1522}}}
\affil[5]{Computational Mathematics, Division of Applied Mathematics, Link\"oping University, Sweden}

\author[6]{Florian~Hindenlang\thanks{\orcid{0000-0002-0439-249X}}}
\affil[6]{Max Planck Institute for Plasma Physics, NMPP division, Garching, Germany}

\author[7]{Gregor~J.~Gassner\thanks{\orcid{0000-0002-1752-1158}}}
\affil[7]{Department of Mathematics and Computer Science, Center for Data and Simulation Science, University of Cologne, Germany}

\date{April 4, 2022}

\makeatletter
\hypersetup{pdfauthor={Hendrik Ranocha et al.}}
\hypersetup{pdftitle={Efficient implementation of modern entropy-based discontinuous Galerkin methods for conservation laws}}
\makeatother

\begin{document}

\maketitle

\begin{abstract}
  \input{abstract.tex}
\end{abstract}


\input{EC_performance.tex}

\section*{Acknowledgments}

\input{acknowledgments.tex}

\printbibliography

\end{document}

%% file: macros.tex
\usepackage{pgfplots}
\usepgfplotslibrary{groupplots}
\pgfplotsset{compat=1.16}
\pgfplotsset{/pgfplots/error bars/error bar style={thick}}
\usetikzlibrary{shapes.geometric}

\definecolor{mycolor1}{RGB}{86,180,233}
\definecolor{mycolor2}{RGB}{0,158,115}
\definecolor{mycolor3}{RGB}{0,114,178}
\definecolor{mycolor4}{RGB}{213,94,0}
\definecolor{mycolor5}{RGB}{204,121,167}
\pgfplotscreateplotcyclelist{modifiedcolor}{%
  color=mycolor1,mark=*\\%
  color=mycolor2,mark=square*\\%
  color=mycolor3,mark options={scale=1.4},mark=triangle*\\%
  color=mycolor4,mark options={scale=1.2},mark=diamond*\\%
  color=mycolor5,mark=pentagon*\\%
}
\pgfplotsset{cycle list name={modifiedcolor}}

\usepackage{xparse}

\let\epsilon\varepsilon
\let\phi\varphi
\let\rho\varrho

\usepackage{xspace}
\newcommand{\cf}[0]{{cf.\@}\xspace}
\newcommand{\eg}[0]{{e.g.\@}\xspace}
\newcommand{\ie}[0]{{i.e.\@}\xspace}

\newcommand{\etc}[0]{{etc.\@}\xspace}
\newcommand{\etal}[0]{{et~al.\@}\xspace}
\newcommand{\trixi}{Trixi.jl\xspace}
\newcommand{\fluxo}{FLUXO\xspace}

\renewcommand{\O}{\mathcal{O}}

\newcommand{\R}{\mathbb{R}}

\renewcommand{\Re}{\operatorname{Re}}
\renewcommand{\Im}{\operatorname{Im}}

\renewcommand{\div}{\operatorname{div}}

\newcommand{\I}{\operatorname{I}}

\newcommand{\diag}{\operatorname{diag}}

\newcommand{\ndims}{\ensuremath{d}}
\newcommand{\nvariables}{\ensuremath{c}}
\newcommand{\ndofs}{\ensuremath{\mathrm{\#DOF}}}
\newcommand{\polydeg}{\ensuremath{p}}
\newcommand{\PID}{\ensuremath{\mathrm{PID}}\xspace}
\newcommand{\normal}{n}

\renewcommand{\vec}[1]{\pmb{#1}}

\newcommand{\Jac}{J}
\newcommand{\covar}{(\Jac a)}
\newcommand{\veccovar}{(\vec{\Jac a})}

\NewDocumentCommand{\opD}{m+g}{%
  \IfNoValueTF{#2}
    {D_{#1}}
    {D_{#1,#2}}%
}
\NewDocumentCommand{\opDsplit}{m+g}{%
  \IfNoValueTF{#2}
    {\widetilde{D}_{#1}}
    {\widetilde{D}_{#1,#2}}%
}
\NewDocumentCommand{\opM}{g}{%
  \IfNoValueTF{#1}
    {M}
    {M_{#1}}%
}
\NewDocumentCommand{\opQ}{g}{%
  \IfNoValueTF{#1}
    {Q}
    {Q_{#1}}%
}
\NewDocumentCommand{\opI}{g}{%
  \IfNoValueTF{#1}
    {I}
    {I_{#1}}%
}
\NewDocumentCommand{\opV}{g}{%
  \IfNoValueTF{#1}
    {V}
    {V_{#1}}%
}
\NewDocumentCommand{\opB}{g}{%
  \IfNoValueTF{#1}
    {B}
    {B_{#1}}%
}
\NewDocumentCommand{\opR}{g}{%
  \IfNoValueTF{#1}
    {R}
    {R_{#1}}%
}
\NewDocumentCommand{\opN}{m+g}{%
  \IfNoValueTF{#2}
    {N_{#1}}
    {N_{#1,#2}}%
}
\NewDocumentCommand{\x}{g}{%
  \IfNoValueTF{#1}
    {\vec{x}}
    {\vec{x}_{#1}}%
}

\NewDocumentCommand{\fnum}{g}{%
  \IfNoValueTF{#1}
    {f^{\mathrm{num}}}
    {f^{\mathrm{num,#1}}}%
}
\NewDocumentCommand{\vecfnum}{g}{%
  \IfNoValueTF{#1}
    {\vec{f}^{\mathrm{num}}}
    {\vec{f}^{\mathrm{num,#1}}}%
}
\NewDocumentCommand{\vecfcorr}{g}{%
  \IfNoValueTF{#1}
    {\vec{f}^{\mathrm{corr}}}
    {\vec{f}^{\mathrm{corr,#1}}}%
}
\NewDocumentCommand{\fvol}{g}{%
  \IfNoValueTF{#1}
    {f^{\smash{\mathrm{vol}}}}
    {f^{\smash{\mathrm{vol,#1}}}}%
}

\newcommand{\VOL}{\vec{\mathrm{VOL}}}

\newcommand{\mean}[1]{{\{\mkern-6mu\{}#1{\}\mkern-6mu\}}}
\newcommand{\prodmean}[2]{{(\mkern-3mu(}#1\,; #2{)\mkern-3mu)}}
\newcommand{\jump}[1]{{[\mkern-3mu[}#1{]\mkern-3mu]}}

\newcommand{\logmean}[1]{\mean{#1}_\mathrm{log}}

%% file: abstract.tex
Many modern discontinuous Galerkin (DG) methods for conservation laws make use
of summation by parts operators and flux differencing to achieve kinetic energy
preservation or entropy stability. While these techniques increase the robustness
of DG methods significantly, they are also computationally more demanding than
standard weak form nodal DG methods. We present several implementation techniques
to improve the efficiency of flux differencing DG methods
that use tensor product quadrilateral or hexahedral elements, in 2D or 3D respectively.
Focus is mostly given to CPUs
and DG methods for the compressible Euler equations, although
these techniques are generally also useful for other physical
systems including the compressible Navier-Stokes and magnetohydrodynamics equations.
We present results using two open source codes, Trixi.jl written in Julia and
FLUXO written in Fortran, to demonstrate that our proposed implementation
techniques are applicable to different code bases and programming languages.

%% file: EC_performance.tex
\section{Introduction}

Stability and robustness of numerical methods are key to efficient and reliable
simulations. However, these properties are not trivial to obtain, in particular
for high-order methods. Usually, techniques to improve the stability and (numerical)
robustness also increase the computational complexity and costs. Thus, special
care must be taken when implementing these methods to avoid losing (too much)
efficiency in practice \cite{maier2021efficient,guermond2022implementation}.

In this article, we focus on so-called flux differencing schemes. These methods
have their origins in the second-order accurate entropy-conserving methods
of Tadmor \cite{tadmor1987numerical,tadmor2003entropy} and have successfully
been extended to high-order methods in periodic \cite{lefloch2002fully} and
bounded domains \cite{fisher2013discretely,ranocha2018comparison,chen2017entropy}.
Flux differencing methods are not only useful for constructing entropy-conservative
or -dissipative (EC/ED) methods but also to recover split forms used, for example,
in finite difference methods \cite{gassner2016split}. While flux differencing
methods are central-type schemes without explicit dissipation, they have been
demonstrated to increase the robustness of numerical methods significantly
\cite{winters2018comparative,klose2020assessing,bergmann2020assessment,rojas2021robustness,sjogreen2017skew,sjogreen2018high,gassner2021novel}.
Thus, these schemes can be used as baseline methods to which dissipation can be
added in a controlled manner \cite{fjordholm2012arbitrarily,flad2017use}.

Flux differencing methods require two key ingredients: a discrete derivative
operator satisfying the summation by parts (SBP) property and a two-point
numerical flux. The SBP property guarantees a discrete equivalent of integration
by parts and allows to transfer properties of the numerical flux to high-order
methods, \eg, entropy conservation, preservation of the kinetic energy
\cite{jameson2008formulation,gassner2014kinetic,chandrashekar2013kinetic,ranocha2020entropy}
or important (quasi-) steady states including pressure equilibria
\cite{ranocha2021preventing} and lake-at-rest-type states for the shallow water
equations \cite{winters2015comparison,ranocha2017shallow,wintermeyer2017entropy}.

SBP operators originate in finite difference methods \cite{kreiss1974finite}.
Due to their attractive properties, they have recently received a lot of interest.
Many common numerical schemes can be formulated in terms of SBP operators,
including finite difference methods \cite{kreiss1974finite,strand1994summation},
finite volume methods \cite{nordstrom2001finite,nordstrom2003finite},
continuous Galerkin methods \cite{hicken2016multidimensional,hicken2020entropy,abgrall2020analysisI},
discontinuous Galerkin (DG) methods \cite{gassner2013skew,carpenter2014entropy,chan2018discretely},
and flux reconstruction methods \cite{huynh2007flux,ranocha2016summation}.
Thus, they can be used to analyze a broad range of numerical methods in a unified fashion
\cite{ranocha2021broad,ranocha2021sbp}. For background information on
SBP operators and further references, we recommend the review articles
\cite{svard2014review,fernandez2014review}.

This article presents our experience with efficient implementation techniques
for flux differencing methods
based on our open source projects
\trixi \cite{ranocha2022adaptive,schlottkelakemper2021purely}
(written in Julia \cite{bezanson2017julia})
and \fluxo\footnote{\url{https://gitlab.com/project-fluxo/fluxo}} (written in Fortran),
e.g., \cite{rueda2021entropy}.
These two codes are written in different languages and use different
approaches in many aspects. In this respect, we posit that the techniques discussed in
this article are general and can be applied successfully to other
(open or closed source) DG codes such as FLEXI \cite{krais2021flexi} and SSDC
\cite{parsani2021ssdc}. Additionally, many of the techniques can also be applied
directly to other SBP schemes including finite difference methods.

We begin our discussion with a brief introduction to SBP methods in
Section~\ref{sec:SBP}. There, we also describe broad techniques for
the efficient implementation of flux differencing schemes applicable
to general conservation laws. Next, we focus on specific aspects related to the
compressible Euler equations in Section~\ref{sec:Euler}. These apply directly
to the advective components of related models such as the compressible Navier-Stokes equations and
magnetohydrodynamics. Having established a solid basis of efficient implementation
techniques, we present some baseline performance benchmarks in
Section~\ref{sec:numerical-experiments-base}.
In Section~\ref{sec:overintegration}, we compare flux differencing to
overintegration, another common technique to increase the robustness of DG methods.
Next, we discuss extensions of the efficient implementation for entropy-based flux differencing DG methods
to general nodal distributions in Section~\ref{sec:Gauss}.
Thereafter, we discuss more invasive optimizations specifically tuned
for the compressible Euler equations and related models in Section~\ref{sec:invasive}.
This discussion includes technical optimizations whose complexity
depends on the programming language.
In Section~\ref{sec:SIMD}, we investigate SIMD (single instruction, multiple data) optimizations and present related performance
improvements.
Finally, we summarize and discuss our results in Section~\ref{sec:summary}.

The code and instructions to reproduce all numerical results shown in this article
are available in our reproducibility repository \cite{ranocha2021efficientRepro}.
We use explicit time integration methods from the Julia library
OrdinaryDiffEq.jl \cite{rackauckas2017differentialequations} for \trixi.

\section{Summation by parts operators and flux differencing}
\label{sec:SBP}

Consider a hyperbolic conservation law
\begin{equation}
\label{eq:hcl}
  \partial_t u(t,x) + \sum\limits_{j=1}^d\partial_j f^j(u(t,x)) = 0,
  \quad t \in (0, T), x \in \Omega,
\end{equation}
where the conserved variables are $u\colon [0,T] \times \Omega \to \Upsilon \subset \R^\nvariables$.
The conservation law \eqref{eq:hcl} must be supplemented by initial and boundary
conditions. Since our techniques for an efficient implementation will concentrate
on volume terms and internal interfaces, we do not present details of boundary
conditions here. Moreover, we will suppress the dependency on time $t \in [0, T]$
and space coordinates $x \in \Omega \subset \R^\ndims$ in the following.

Entropy estimates for conservation laws are based on the chain rule and symmetry
properties of the differential operator with respect to the $L^2$ scalar product.
Thus, these two ingredients are mimicked discretely by two-point numerical fluxes
and appropriate SBP derivative operators. We consider the method of lines and
first discretize \eqref{eq:hcl} in space. The spatial semidiscretization
uses a division of the domain $\Omega$ into non-overlapping elements $\Omega_l$.
The numerical solution is represented by a vector in a finite dimensional space
on each element. For example, DG methods typically use polynomial spaces \cite{hesthaven2007nodal,kopriva2009implementing}. On
each element, SBP operators are applied.

\begin{definition}
  A $\polydeg$-th order accurate \emph{derivative matrix} $\opD{j}$ satisfies
  $
    \forall k \in \{0, \dots, \polydeg\}\colon
      \opD{j} \x{j}^k = k \x{j}^{k-1},
  $
  with the convention $\x^0 = \vec{1}$. We say $\opD{j}$ is consistent if $p \ge 0$.
\end{definition}

\begin{definition}
  A (first-derivative) \emph{SBP operator} on a $d$-dimensional element $\Omega_l$
  consists of
  consistent first-derivative matrices $\opD{j}$, $j \in \{1, \dots, \ndims\}$,
  a symmetric and positive-definite matrix $\opM$ approximating the scalar product
  on $L^2(\Omega_l)$,
  a restriction operator $\opR$ approximating the restriction of functions on the
  volume $\Omega_l$ to the boundary $\partial \Omega_l$,
  a symmetric and positive-definite matrix $\opB$ approximating the scalar product
  on $L^2(\partial \Omega_l)$,
  and multiplication operators $\opN{j}$ representing the multiplication of functions
  on $\partial \Omega_l$ by the $j$-th component of the outer unit normal
  $\normal$ such that
  \begin{align}
  \label{eq:SBP}
    \opM \opD{j} + \opD{j}^T \opM & = \opR^T \opB \opN{j} \opR.
  \end{align}
  We refer to $\opM$ as a mass matrix or norm matrix\footnote{The term ``mass matrix''
  is common for finite element methods. In the finite difference SBP community,
  the name ``norm matrix'' is more common.}.
\end{definition}

\begin{example}
  We mainly focus on tensor product elements using nodal Legendre-Gauss-Lobatto (LGL)
  bases since they are SBP operators with diagonal mass matrix $\opM$ including
  the boundary nodes \cite{gassner2013skew}. In particular, the boundary operators
  $\opR^T \opB \opN{j} \opR$
  are diagonal. The resulting method is often called discontinuous Galerkin
  spectral element method (LGL-DGSEM).
\end{example}

Interface coupling between elements is usually performed via numerical fluxes
in DG methods. We write normal vectors as $\normal \in \mathbb{S}^{\ndims-1}$,
where $\mathbb{S}^{\ndims-1}$ is the unit sphere in $\R^{\ndims}$.

\begin{definition}
\label{def:numerical-fluxes}
  A (Cartesian) \emph{numerical flux} in the $j$-th coordinate direction is a
  Lipschitz continuous mapping $\fnum{j} \colon \Upsilon^2 \to \R^\nvariables$ satisfying
  $\forall u \in \Upsilon \colon \fnum{j}(u, u) = f^j(u)$. It is \emph{symmetric}
  if $\forall u_1, u_2 \in \Upsilon \colon \fnum{j}(u_1, u_2) = \fnum_{j}(u_2, u_1)$.
  A directional numerical flux is a Lipschitz continuous mapping
  $\fnum \colon \Upsilon^2 \times \mathbb{S}^{\ndims-1} \to \R^\nvariables$ satisfying
  $\forall u \in \Upsilon, \normal \in \mathbb{S}^{\ndims-1} \colon \fnum(u, u, \normal) = \sum_{j=1}^\ndims \normal_j f^j(u)$
  and
  $\forall u_{\mathrm{in}}, u_{\mathrm{out}} \in \Upsilon, \normal \in \mathbb{S}^{\ndims-1} \colon \fnum(u_{\mathrm{in}}, u_{\mathrm{out}}, \normal) = -\fnum(u_{\mathrm{out}}, u_{\mathrm{in}}, -\normal)$.
  It is \emph{symmetric} if
  $\forall u_1, u_2 \in \Upsilon, \normal \in \mathbb{S}^{\ndims-1} \colon \fnum(u_1, u_2, \normal) = \fnum(u_2, u_1, \normal)$.
\end{definition}

We assume that the conservation law \eqref{eq:hcl} is equipped with an entropy function $U$ and
corresponding entropy fluxes $F^j$ satisfying $\partial_u U \cdot \partial_u f^j = \partial_u F^j$.
We denote the \emph{entropy variables} as $w = \partial_u U$ and the
\emph{flux potentials} as $\psi^j = w \cdot f^j - F^j$. Then, EC/ED fluxes are
given as follows \cite{tadmor1987numerical}.
\begin{definition}
  A Cartesian numerical flux is EC if
  $\forall u_-, u_+ \in \Upsilon \colon
    (w(u_+) - w(u_-)) \cdot \fnum{j}(u_-, u_+) - (\psi^j(u_+) - \psi^j(u_-)) = 0$
  and ED if $\le$ holds instead of equality.
  A directional numerical flux is EC if
  $\forall u_{\mathrm{in}}, u_{\mathrm{out}} \in \Upsilon,
    \normal \in \mathbb{S}^{\ndims-1} \colon
    (w(u_{\mathrm{out}}) - w(u_{\mathrm{in}})) \cdot \fnum(u_{\mathrm{in}}, u_{\mathrm{out}}, \normal) - \sum_{j=1}^\ndims \normal_j (\psi^j(u_{\mathrm{out}}) - \psi^j(u_{\mathrm{in}})) = 0$
  and ED if $\le$ holds instead of equality.
\end{definition}

Collecting numerical fluxes at the interface between a given element and its
neighbors in $\vecfnum$, a typical strong form DG formulation on one element
can be written as
\begin{equation}
\label{eq:strong-form-SBP}
  \partial_t \vec{u}
  + \sum_{j=1}^\ndims \opD{j} \vec{f}^j
  + \opM^{-1} \opR^T \opB \left( \vecfnum - \sum_{j=1}^\ndims \opN{j} \opR \vec{f}^j \right)
  = \vec{0}.
\end{equation}
This mimicks, for all polynomials $v$ of degree $\polydeg$, the variational form of \eqref{eq:hcl}
\begin{equation}
  \int_{\Omega_l} v \cdot \partial_t u
  + \int_{\Omega_l} v \cdot \partial_j f^j
  + \int_{\partial \Omega_l} \left( \fnum - \sum_{j=1}^\ndims \normal_j f^j \right)
  = 0.
\end{equation}
Flux differencing methods replace the volume term $\sum_{j=1}^\ndims \opD{j} \vec{f}^j$ by
another volume term $\VOL$ using so-called volume fluxes $\fvol$ \cite{gassner2016split}, which are
\emph{symmetric} numerical fluxes, resulting in
\begin{equation}
\label{eq:fluxdiff-SBP}
  \partial_t \vec{u}
  + \VOL
  + \opM^{-1} \opR^T \opB \left( \vecfnum - \sum_{j=1}^\ndims \opN{j} \opR \vec{f}^j \right)
  = \vec{0},
  \quad
  \VOL_i = \sum_{j=1}^\ndims \sum_k 2 (\opD{j})_{i,k} \fvol{j}(\vec{u}_i, \vec{u}_k).
\end{equation}
Here, the sum $\sum_k$ is performed over all degrees of freedom on the given
element. In the following, we assume that the mass matrix $\opM$ and the
boundary operators $\opR^T \opB \opN{j} \opR$ are diagonal. Then, this flux
differencing form can be rewritten in locally conservative form; it is EC/ED if
the volume fluxes are EC and the surface fluxes are EC/ED \cite{carpenter2014entropy,gassner2016split}.
Moreover, it is of the same order of accuracy as  the derivative operator for
general symmetric volume fluxes \cite{ranocha2018comparison,crean2018entropy}.

On Legendre-Gauss-Lobatto tensor product elements with polynomials of degree $\polydeg$,
the volume terms of the flux differencing form \eqref{eq:fluxdiff-SBP} require
asymptotically $(\polydeg + 1)$-times more flux evaluations than the classical
strong form \eqref{eq:strong-form-SBP}, see Table~\ref{tab:computational-complexity}.
However, the asymptotic number of floating point operations without computing
fluxes is the same for both volume terms (basically one matrix vector product
along each line in a tensor product layout per spatial dimension).

\begin{table}[!htb]
\centering
\caption{Computational complexity of strong form and flux differencing volume
         terms on Legendre-Gauss-Lobatto tensor product elements in $\ndims$ spatial
         dimensions using polynomials of degree $\polydeg$. Pointwise fluxes
         are used in the strong form, while two-point fluxes are used in the flux differencing
         formulation.}
\label{tab:computational-complexity}
\begin{tabular*}{\linewidth}{@{\extracolsep{\fill}} lll @{}}
  \toprule
  & Strong form \eqref{eq:strong-form-SBP} &
    Flux differencing \eqref{eq:fluxdiff-SBP}
  \\
  \midrule
  Flux evaluations &
    $\ndims (\polydeg + 1)^\ndims$ &
    $\O\bigl(  \ndims (\polydeg + 1)^{\ndims + 1} \bigr)$
  \\
  Floating point operations (without fluxes) &
    $\O\bigl( \ndims (\polydeg + 1)^{\ndims + 1} \bigr)$ &
    $\O\bigl( \ndims (\polydeg + 1)^{\ndims + 1} \bigr)$
  \\
  \bottomrule
\end{tabular*}
\end{table}

Having introduced the basic form of flux difference semidiscretizations
\eqref{eq:fluxdiff-SBP}, we next present techniques for their efficient
implementation. Most of these implementation aspects are designed to reduce the total
number of operations, based on the hypothesis that the evaluation of fluxes $f^j$
and volume fluxes $\fvol$ is expensive, which holds for the compressible Euler
equations and related models.

\subsection{Separation of volume and surface terms}
\label{sec:separation-volume-surface}

Using the SBP property \eqref{eq:SBP}, the strong form \eqref{eq:strong-form-SBP}
is equivalent to the classical weak form semidiscretization \cite{kopriva2010quadrature}
\begin{equation}
\label{eq:weak-form-SBP}
  \partial_t \vec{u}
  - \sum_{j=1}^\ndims \opM^{-1} \opD{j}^T \opM \vec{f}^j
  + \opM^{-1} \opR^T \opB \vecfnum
  = \vec{0}.
\end{equation}
The weak form \eqref{eq:weak-form-SBP} is slightly more computationally attractive as no
additional pointwise flux evaluations are required for the surface terms. The
same efficiency optimization can be achieved for flux differencing discretizations
by rewriting the volume terms in \eqref{eq:fluxdiff-SBP} to use the
\emph{flux differencing operator}
\begin{equation}
\label{eq:opDsplit}
  \opDsplit{j} = 2 \opD{j} - \opM^{-1} \opR^T \opB \opN{j} \opR,
\end{equation}
which is skew-symmetric with respect to $\opM$, see Section~\ref{sec:symmetry}
below.
Indeed, since we assume that the mass matrix and the boundary operators are
diagonal, $\opM^{-1} \opR^T \opB \opN{j} \opR$ is also diagonal. Thus,
\begin{equation}
\label{eq:split-form-vol-term}
\begin{aligned}
  \sum_k (\opDsplit{j})_{i,k} \fvol{j}(\vec{u}_i, \vec{u}_k)
  &=
  \sum_k 2 (\opD{j})_{i,k} \fvol{j}(\vec{u}_i, \vec{u}_k)
  - (\opM^{-1} \opR^T \opB \opN{j} \opR)_{i,i} \fvol{j}(\vec{u}_i, \vec{u}_i).
\end{aligned}
\end{equation}
From the consistency of the volume flux $\fvol{j}$, the last term in \eqref{eq:split-form-vol-term}
simplifies to be
\begin{equation}
  \left( \opM^{-1} \opR^T \opB \opN{j} \opR \vec{f}^j \right)_i,
\end{equation}
which is exactly the surface flux term in \eqref{eq:fluxdiff-SBP}. Hence, an
optimized version of \eqref{eq:fluxdiff-SBP} that uses the flux differencing operators
$\opDsplit{j}$ \eqref{eq:opDsplit} takes the form
\begin{equation}
\label{eq:fluxdiff-SBP-opDsplit}
  \partial_t \vec{u}
  + \VOL
  + \opM^{-1} \opR^T \opB \vecfnum
  = \vec{0},
  \quad
  \VOL_i = \sum_{j=1}^\ndims \sum_k (\opDsplit{j})_{i,k} \fvol{j}(\vec{u}_i, \vec{u}_k).
\end{equation}

\subsection{Symmetry properties of numerical fluxes and SBP operators}
\label{sec:symmetry}

The flux differencing matrix $\opDsplit{j}$ \eqref{eq:opDsplit} is skew-symmetric
with respect to $\opM$. Indeed, the SBP property \eqref{eq:SBP} yields
\begin{equation}
  \opM \opDsplit{j}
  =
  2 \opM \opD{j} - \opR^T \opB \opN{j} \opR
  =
  - 2 \opD{j}^T \opM + \opR^T \opN{j}^T \opB \opR
  =
  - (\opM \opDsplit{j})^T.
\end{equation}
In particular, the diagonal entries $(\opDsplit{j})_{i,i}$ are zero. Thus,
exploiting the symmetry of the volume fluxes,
it suffices to compute two-point volume fluxes $\fvol{j}(\vec{u}_i, \vec{u}_k)$
only for combinations with, say, $i < k$. This saves more than half of the total
number of volume flux evaluations.

\subsection{Sparsity structure of tensor product operators and curvilinear coordinates}
\label{sec:directional-fluxes}

Given one-dimensional SBP operators $\opD{j}{1D}$, two-dimensional SBP operators
on the tensor product domain can be constructed as $\opD{1} = \opD{1}{1D} \otimes \I$
and $\opD{2} = \I \otimes \opD{2}{1D}$. Thus, they are naturally sparse. Exploiting
this sparsity structure is a fundamental step in an efficient implementation.
This is easily possible on a Cartesian mesh using the flux differencing form
\eqref{eq:fluxdiff-SBP} directly with Cartesian volume fluxes. Here, we describe
how to transfer the same efficiency to curvilinear meshes.

We restrict this section to two spatial dimensions to make the presentation simpler.
Assume we are given a Cartesian reference quadrilateral with coordinates $\xi^i$ and a mapped
curved version with coordinates $x^i$. The discrete Jacobian in 2D can be calculated
directly as
\begin{equation}
  \vec{\Jac} = \veccovar^2_2 \, \veccovar^1_1  -  \veccovar^1_2 \, \veccovar^2_1,
\end{equation}
where
\begin{equation}
  \veccovar^1_1 =  \diag\bigl(\opD{2}{\xi} \vec{x^2}\bigr), \quad
  \veccovar^1_2 = -\diag\bigl(\opD{2}{\xi} \vec{x^1}\bigr), \quad
  \veccovar^2_1 = -\diag\bigl(\opD{1}{\xi} \vec{x^2}\bigr), \quad
  \veccovar^2_2 =  \diag\bigl(\opD{1}{\xi} \vec{x^1}\bigr), \quad
\end{equation}
are the components of scaled contravariant basis vectors $\covar^n_j$ in 2D \cite[Chapter~6]{kopriva2009implementing},
$\vec{x^i}$ is the vector containing the nodal values of the curvilinear coordinates
of an element, and
$\opD{j}{\xi}$ are tensor product SBP operators on the Cartesian reference coordinates $\xi^j$.
Using these ingredients, SBP operators on the curved element can be constructed as
\cite{alund2019encapsulated}
\begin{equation}
\label{eq:curved-SBP-operators}
\opD{j}{x}=
  \frac{1}{2} \vec{\Jac}^{-1} \sum_{n=1}^\ndims \bigl(
      \veccovar^n_j \opD{n}{\xi} + \opD{n}{\xi} \veccovar^n_j  \bigr), \qquad 1\leq j\leq\ndims.
\end{equation}
The curvilinear SBP operators \eqref{eq:curved-SBP-operators} in the
volume terms of the flux differencing form \eqref{eq:fluxdiff-SBP} are linear
combinations of the underlying Cartesian SBP operators. Specifically,
the entry corresponding to nodes $i$ and $k$ is
\begin{equation}
  (\opD{j})_{i,k}
  =
  \frac{1}{2 \Jac_{i}} \sum_{n=1}^\ndims \biggl(
    \sum_l \bigl(\veccovar^n_j\bigr)_{i,l} \bigl(\opD{n}{\xi}\bigr)_{l,k}
    + \sum_l \bigl(\opD{n}{\xi}\bigr)_{i,l} \bigl(\veccovar^n_j\bigr)_{l,k}
  \biggr),
\end{equation}
where we used that $\vec{\Jac}$ is diagonal. Since $\veccovar^n_j$ is diagonal,
too,
\begin{equation}
  (\opD{j})_{i,k}
  =
  \frac{1}{2 \Jac_{i}} \sum_{n=1}^\ndims \Bigl(
    \bigl(\veccovar^n_j\bigr)_{i,i} \bigl(\opD{n}{\xi}\bigr)_{i,k}
    + \bigl(\opD{n}{\xi}\bigr)_{i,k} \bigl(\veccovar^n_j\bigr)_{k,k}
  \Bigr).
\end{equation}
This can be abbreviated as
\begin{equation}
  (\opD{j})_{i,k} = \frac{1}{\Jac_{i}}\sum_{n=1}^\ndims \alpha^{j,n}_{i,k} (\opD{n}{\xi})_{i,k}, \qquad 1\leq j\leq\ndims,
\end{equation}
where $\alpha^{j,n}_{i,k}=\frac{1}{2} \bigl((\covar^n_j)_{i}+ (\covar^n_j)_{k}  \bigr)$,
which is the arithmetic average of the scaled contravariant basis vectors at nodes $i$ and $k$.
Thus, the volume terms are recast to be
\begin{equation}
\label{eq:curvilinear-rewrite}
  \sum_{j=1}^\ndims \sum_k 2 (\opD{j})_{i,k} \fvol{j}(\vec{u}_i, \vec{u}_k)
  =
  \frac{1}{J_{i}} \sum_{j=1}^\ndims \sum_k 2
    \biggl( \sum_{n=1}^\ndims \alpha^{j,n}_{i,k} (\opD{n}{\xi})_{i,k} \biggr)
    \fvol{j}(\vec{u}_i, \vec{u}_k).
\end{equation}
If implemented in this form, we need only compute Cartesian volume fluxes instead
of directional ones, which is usually slightly more efficient. However, the form given in \eqref{eq:curvilinear-rewrite}
loses the sparsity of the individual tensor product operators. Thus,
it is advantageous to rearrange terms as
\begin{equation}
\label{eq:curvilinear-rearrange}
  \sum_{j=1}^\ndims \sum_k 2
    \biggl( \sum_{n=1}^\ndims \alpha^{j,n}_{i,k} (\opD{n}{\xi})_{i,k} \biggr)
    \fvol{j}(\vec{u}_i, \vec{u}_k)
  =
  \sum_{n=1}^\ndims \sum_k 2 (\opD{n}{\xi})_{i,k}
    \biggl( \sum_{j=1}^\ndims \alpha^{j,n}_{i,k} \fvol{j}(\vec{u}_i, \vec{u}_k) \biggr).
\end{equation}
The sum over the Cartesian volume fluxes is effectively a directional volume
flux $\fvol(\vec{u}_i, \vec{u}_k, \alpha^{\cdot, n}_{i, k})$ in the direction
of the arithmetic average of the contravariant basis vectors at nodes $i$ and $k$.

In this form, the sparsity structure of the underlying Cartesian tensor product
SBP operators is completely retained. Since the evaluation of volume fluxes is
usually relatively expensive and the evaluation of directional volume fluxes is
only marginally more expensive than the computation of their Cartesian equivalents,
it is advantageous to use the latter form \eqref{eq:curvilinear-rearrange}.
Of course, the usual care must be
taken to ensure free-stream preservation \etc in multiple space dimensions when
computing the contravariant basis vectors \cite{kopriva2006metric,thomas1979geometric}. Moreover,
the techniques discussed in the previous Sections~\ref{sec:separation-volume-surface}
and \ref{sec:symmetry} can still be applied.

\subsection{Discussion}

The impact of the optimizations discussed in this section depend on the
complexity of the numerical fluxes. For relatively expensive volume fluxes,
such as EC fluxes for the compressible Euler equations, taking
advantage of the symmetry (Section~\ref{sec:symmetry}) and sparsity
(Section~\ref{sec:directional-fluxes}) properties are most important.
These optimizations should always be applied before focusing on problem-specific
optimizations like the ones discussed in the following section.
On top of these optimizations, common efficient implementation techniques for
discontinuous Galerkin methods still apply. For example, it is usually
efficient to compute the numerical fluxes at surfaces once per surface (instead
of once per element) and use them to update the right-hand side on CPUs.

\section{Compressible Euler equations}
\label{sec:Euler}

To discuss further efficient implementation strategies we consider the compressible Euler equations
\begin{equation}
\label{eq:Euler}
  \partial_t
  \begin{pmatrix}
    \rho \\
    \rho v_i \\
    \rho e
  \end{pmatrix}
  + \sum_{j=1}^\ndims \partial_j
  \begin{pmatrix}
    \rho v_j \\
    \rho v_j v_i + p \delta_{ij} \\
    (\rho e + p) v_j
  \end{pmatrix}
  =
  0,
  \qquad 1 \leq i \leq \ndims,
\end{equation}
as an example system of conservation laws.
Here, $\rho$ is the fluid density, $v$ the velocity, $e$ the specific total
energy, and $p$ the pressure. We assume a perfect gas law with ratio of specific
heats $\gamma$, \ie,
\begin{equation}
  p = (\gamma - 1) \left( \rho e - \frac{1}{2} \rho |v|^2 \right).
\end{equation}

\subsection{Logarithmic mean}
\label{sec:logmean}

Since the seminal work of \citet{ismail2009affordable} on affordable EC fluxes
for the compressible Euler equations, the logarithmic mean
\begin{equation}
\label{eq:logmean}
  \logmean{a} = \jump{a} / \jump{\log(a)}
\end{equation}
has played a crucial role. Indeed, it is even necessary if some desirable additional
properties are to be satisfied \cite{ranocha2021preventing}. Here, we have used
the common jump notation
\begin{equation}
  \jump{a} = a_+ - a_-.
\end{equation}

Since a naive implementation of the logarithmic mean \eqref{eq:logmean} is
subject to floating point accuracy issues, special care must be taken when
$a_+ \approx a_-$. Here, we present an efficient version based on
Algorithm~\ref{alg:logmean-Ismail-Roe} presented in \cite{ismail2009affordable}.

\begin{algorithm}[!hbt]
  \caption{Computation of the logarithmic mean as described by
           \citet{ismail2009affordable}.}
  \label{alg:logmean-Ismail-Roe}
  \begin{algorithmic}
    \Require Input $a_-, a_+ > 0$
    \Ensure Stable approximation of the logarithmic mean $\logmean{a}$
    \State $\xi = a_- / a_+$
    \State $f = (\xi - 1) / (\xi + 1)$
    \State $u = f \cdot f$
    \State $\epsilon = 10^{-4}$ \Comment{for 64 bit floating point numbers}
    \If{$u < \epsilon$}
      \State $F = 1 + u / 3 + u \cdot u / 5 + u \cdot u \cdot u / 7$
    \Else
      \State $F = (\log(\xi) / 2) / f$
    \EndIf
    \State \Return $(a_- + a_+) / (2 F)$
  \end{algorithmic}
\end{algorithm}

Since divisions are more expensive (in terms of latency and inverse throughput)
than multiplications on modern (CPU) hardware \cite{fog2021instruction}, this
algorithm can be improved by re-writing divisions in terms of cheaper multiplications,
even if more additions are required on top; these can be combined into fused
multiply-add (FMA) instructions. The resulting optimized implementation of the
logarithmic mean is given in Algorithm~\ref{alg:logmean-optimized}.

\begin{algorithm}[!htb]
  \caption{Optimized computation of the logarithmic mean.}
  \label{alg:logmean-optimized}
  \begin{algorithmic}
    \Require Input $a_-, a_+ > 0$
    \Ensure Stable approximation of the logarithmic mean $\logmean{a}$
    \State $u = (a_- \cdot (a_- - 2 a_+) + a_+ \cdot a_+) /
                (a_- \cdot (a_- + 2 a_+) + a_+ \cdot a_+)$
                \Comment{equivalent to $f^2, f = \frac{\xi - 1}{\xi + 1}, \xi = \frac{a_-}{a_+}$}
    \State $\epsilon = 10^{-4}$
    \If{$u < \epsilon$}
      \State\Return $(a_- + a_+) /
                     \bigl( 2 + u \cdot (2/3 + u \cdot (2/5 + u \cdot 2/7)) \bigr)$
                      \Comment{use Horner's rule}
    \Else
      \State \Return $(a_+ - a_-) / \log(a_+ / a_-)$
    \EndIf
  \end{algorithmic}
\end{algorithm}

Another variant of Algorithm~\ref{alg:logmean-optimized} replaces the
division in the ``if'' branch by a multiplication using a polynomial approximation
of $1 / \log(\cdot)$. For example, the term
\begin{equation}
  (a_- + a_+) / \bigl( 2 + u \cdot (2/3 + u \cdot (2/5 + u \cdot 2/7)) \bigr)
\end{equation}
can be replaced by
\begin{equation}
  (a_- + a_+) \cdot \bigl( 1/2 + u \cdot ((-1/6) + u \cdot ((-2/45) + u \cdot (-22/945))) \bigr)
\end{equation}
without changing other parts of Algorithm~\ref{alg:logmean-optimized}. Our benchmarks
indicate that this does not result in significant performance differences on
the hardware currently available to us.

Some EC numerical fluxes for the compressible Euler equations presented
in the following sections contain also the inverse logarithmic mean, \ie, factors
of the form $1 / \logmean{a}$. Since the computation of the logarithmic mean in
Algorithm~\ref{alg:logmean-optimized} already contains a division in the last
step, it is advantageous to avoid the additional division and also implement
Algorithm~\ref{alg:inv_logmean-optimized} for computing the inverse logarithmic
mean.

\begin{algorithm}[!htb]
  \caption{Optimized computation of the inverse logarithmic mean.}
  \label{alg:inv_logmean-optimized}
  \begin{algorithmic}
    \Require Input $a_-, a_+ > 0$
    \Ensure Stable approximation of the inverse logarithmic mean $1 / \logmean{a}$
    \State $u = (a_- \cdot (a_- - 2 a_+) + a_+ \cdot a_+) /
                (a_- \cdot (a_- + 2 a_+) + a_+ \cdot a_+)$
                \Comment{equivalent to $f^2, f = \frac{\xi - 1}{\xi + 1}, \xi = \frac{a_-}{a_+}$}
    \State $\epsilon = 10^{-4}$
    \If{$u < \epsilon$}
      \State \Return $\bigl( 2 + u \cdot (2/3 + u \cdot (2/5 + u \cdot 2/7)) \bigr) /
                      (a_- + a_+)$
                      \Comment{use Horner's rule}
    \Else
      \State \Return $\log(a_+ / a_-) / (a_+ - a_-)$
    \EndIf
  \end{algorithmic}
\end{algorithm}

\subsection{Numerical fluxes for the compressible Euler equations}
\label{sec:numflux-Euler}

As described in Section~\ref{sec:directional-fluxes}, it is advantageous to use
directional volume fluxes for flux differencing on non-Cartesian meshes. First,
we present such a directional form of the non-EC flux of Shima \etal \cite{shima2021preventing},
which is kinetic energy and pressure equilibrium preserving.
\begin{equation}
\label{eq:flux_shima_etal}
\begin{aligned}
  f_{\rho} &= \mean{\rho} \mean{v \cdot \normal},
  \\
  f_{\rho v} &= f_{\rho} \mean{v} + \mean{p} \normal,
  \\
  f_{\rho e} &= f_{\rho} \frac{\prodmean{v}{v}}{2}
                + \mean{p} \mean{v \cdot \normal} \frac{1}{\gamma - 1}
                + \prodmean{p}{v \cdot \normal},
\end{aligned}
\end{equation}
where we introduced the arithmetic mean
\begin{equation}
  \mean{a} = \frac{a_+ + a_-}{2}
\end{equation}
and the product mean
\begin{equation}
\label{eq:prodmean}
  \prodmean{a }{ b}
  = \frac{a_+ \cdot b_- + a_- \cdot b_+}{2}
  = 2 \mean{a} \cdot \mean{b} - \mean{a \cdot b}.
\end{equation}
The version \eqref{eq:flux_shima_etal} already clarifies how common subexpressions
can be used efficiently. In the spirit of Section~\ref{sec:logmean}, modern
hardware will benefit from avoiding divisions by storing $1 / (\gamma - 1)$ and
turning the division by $\gamma - 1$ into a multiplication.

Next, we present a directional version of the EC flux of Ranocha \cite{ranocha2018comparison,ranocha2018thesis,ranocha2020entropy},
which is also kinetic energy and pressure equilibrium preserving.
\begin{equation}
\label{eq:flux_ranocha}
\begin{aligned}
  f_{\rho} &= \logmean{\rho} \mean{v \cdot \normal},
  \\
  f_{\rho v} &= f_{\rho} \mean{v} + \mean{p} \normal,
  \\
  f_{\rho e} &= f_{\rho} \frac{\prodmean{v}{v}}{2}
                + f_{\rho} \frac{1}{\logmean{\rho / p}} \frac{1}{\gamma - 1}
                + \prodmean{p }{v \cdot \normal}.
\end{aligned}
\end{equation}
The direct and inverse logarithmic means should use Algorithms~\ref{alg:logmean-optimized} and \ref{alg:inv_logmean-optimized}, respectively.
Moreover, the
inverse logarithmic mean in \eqref{eq:flux_ranocha} can also be rewritten as
\begin{equation}
  \frac{1}{\logmean{\rho / p}}
  =
  \frac{\log\bigl( (\rho_+ / p_+) / (\rho_- / p_-) \bigr)}
       {\rho_+ / p_+ - \rho_- / p_-}
  =
  p_+ p_- \frac{\log\bigl( (\rho_+ p_-) / (\rho_- p_+) \bigr)}
               {\rho_+ p_- - \rho_- p_+}
  =
  p_+ p_- \frac{1}{\logmean{\rho_+ p_-, \rho_- p_+}}
\end{equation}
to further avoid divisions, resulting in a speed-up on modern hardware.
Nevertheless, EC fluxes such as \eqref{eq:flux_ranocha} involving the logarithmic
mean are computationally more demanding than fluxes using only the arithmetic
mean such as \eqref{eq:flux_shima_etal}. Thus, we will sometimes refer to them
as ``cheap'' (flux of Shima \etal) and ``expensive'' (flux of Ranocha) fluxes.
The same optimizations described here can also be applied successfully to
other entropy-conserving two-point fluxes, e.g., the ones of Ismail and Roe
\cite{ismail2009affordable} or Chandrashekar \cite{chandrashekar2013kinetic}.

\section{Numerical experiments}
\label{sec:numerical-experiments-base}

Here, we present numerical experiments and benchmarks ranging from
microbenchmarks of single numerical flux evaluations to full simulation runs.
The code and instructions to reproduce all numerical results shown in this article
are available in our reproducibility repository \cite{ranocha2021efficientRepro}.
The benchmark results shown in this paper were obtained on a dual socket compute
node with two \SI{2.5}{GHz} Intel\textregistered\ Xeon\textregistered\ Gold 6248 20-core processors and \SI{384}{GiB} RAM,
except where noted otherwise.
We only report serial performance using one core; neither shared memory parallelization
(multiple threads, e.g., \texttt{OpenMP}) nor distributed memory parallelization
(e.g., \texttt{MPI}) is used.
The \trixi code was executed using version v1.7.0 of Julia \cite{bezanson2017julia}
with bound checking disabled and otherwise default options of the official
binaries.
\fluxo was compiled in \texttt{Release} mode with the
Intel\textregistered\ Fortran Compiler 19.1.3, i.e., using the flags
\texttt{-O3}, \texttt{-xHost}, \texttt{-shared-intel}, \texttt{-inline-max-size=1500},
\texttt{-no-inline-max-total-size}, and \texttt{-no-prec-div},
which generates code optimized for the current processor architecture.
Note that we are at an optimization level where the particular instruction
set used can make a noticeable difference in the performance. For example,
restricting the Fortran compiler to AVX2 reduces the performance index
described below by up to \SI{19}{\percent}.
Since the preferred vector width for the LLVM version shipped with Julia
v1.7 is 256 bits, all numerical experiments conducted with \trixi only use the
AVX2 instruction set.

\subsection{Baseline performance results on Cartesian and curved meshes}
\label{sec:Cartesian-vs-curved}

Meshes and numerical algorithms for PDEs support different
geometric features such as curved coordinates, nonconforming interfaces, and
an unstructured connectivity. Sacrificing some of these features can increase
the performance and reduce code complexity. Here, we compare different mesh
types available in \trixi and the standard mesh type of \fluxo to provide
guidance as to whether practitioners might choose to drop certain geometric
features for performance if the problem setup allows it. Specifically, we
compare the serial performance on
\begin{itemize}
  \item the \codeinline{TreeMesh},
        a Cartesian, $h$-nonconforming, tree-structured mesh of \trixi,
  \item the \codeinline{StructuredMesh},
        a curved, conforming, structured mesh of \trixi,
  \item the \codeinline{P4estMesh},
        a curved, $h$-nonconforming, unstructured mesh of \trixi,
  \item the three-dimensional, curved, $h$-nonconforming, unstructured mesh of \fluxo.
\end{itemize}

\trixi and \fluxo implement all efficient implementation techniques discussed
in Sections~\ref{sec:SBP} and \ref{sec:Euler}. On top of that,
for the compressible Euler equations
\fluxo precomputes primitive variables, $1 / \rho$, and
$|v|^2$ at all nodes before computing the volume flux terms. Such an invasive
optimization improves the performance further, as discussed in
Section~\ref{sec:primitive_variables}. Furthermore, both \trixi and \fluxo
use explicit inlining of volume fluxes (and setting the polynomial degree at
compile time for \fluxo), as discussed in Section~\ref{sec:compile-time}.
Since \fluxo is natively a 3D code, 2D results will only be shown for \trixi.

For these performance benchmarks, we use the isentropic vortex setup, a widely
used benchmark problem \cite{shu1997essentially} with an analytical solution.
We use slightly different parameters than the original setup to make sure that
all possible paths in the logarithmic mean are followed. In particular, we
strongly increase the strength of the vortex to $\epsilon = 20$. Hence, the
initial conditions read as
\begin{equation}
\label{eq:isentropic-vortex}
\begin{gathered}
  T = T_0 - \frac{(\gamma-1) \epsilon^2}{8 \gamma \pi^2} \exp\bigl(1-r^2\bigr),
  \quad
  \rho = \rho_0 (T / T_0)^{1 / (\gamma - 1)},
  \\
  v = v_0 + \frac{\varepsilon}{2 \pi} \exp\bigl((1-r^2) / 2\bigr) (-x_2, x_1, 0)^T,
\end{gathered}
\end{equation}
in 3D and its 2D analog, where $r$ is the distance from the origin,
$T = p / \rho$ the temperature, $\rho_0 = 1$ the background density, $v_0 = (1, 1, 0)^T$ the background velocity,
$p_0 = 10$ the background pressure, $\gamma = 1.4$, and $T_0 = p_0 / \rho_0$ the background
temperature. The domain $[-5, 5]^\ndims$ is equipped with periodic boundary conditions.

While all meshes allow more features, we restrict them to a simple Cartesian,
conforming, structured setup here. We use eight elements per coordinate direction.
The LGL-DGSEM discretizations use flux differencing with the same numerical flux in the
volume and at the interfaces. The spatial semidiscretization is integrated in \trixi for $50$
time steps using the nine-stage, fourth-order FSAL Runge-Kutta method of
\cite{ranocha2021optimized} with error-based step size control using a tolerance
of $10^{-8}$. In \fluxo, we integrate in time for $90$ time steps using the five-stage, fourth-order
Runge-Kutta method of \cite{carpenter1994fourth} with CFL-based step size control.
This ensures that both codes use the same number of right-hand side (RHS) evaluations.

For baseline performance benchmarks, we use the performance index \PID as the measure
of runtime efficiency.
\begin{definition}
  The \emph{performance index} \PID is defined as the runtime of one evaluation
  of the spatial semidiscretization divided by the total number of degrees of
  freedom $\ndofs$. Here and in the following, each DG volume node counts as a
  single DOF, i.e., $\ndofs$ is given by $(\polydeg + 1)^\ndims$ times the number
  of elements.
\end{definition}
That is, a \emph{lower} \PID means a \emph{faster} execution of the code.
For all \PID benchmarks, we average the wall clock time per RHS evaluation over
a complete simulation and take the mean value (and standard deviation) of five
runs.

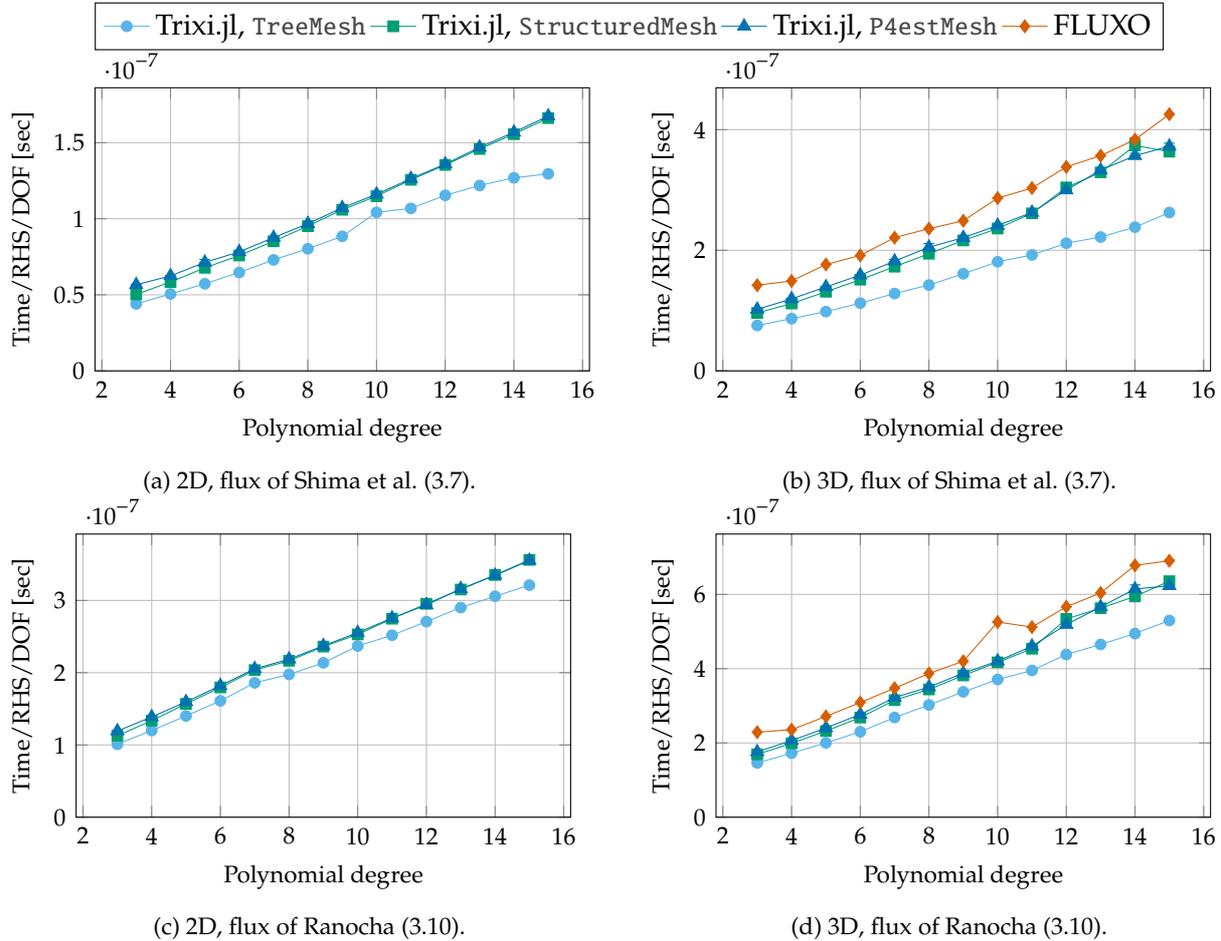
\begin{figure}[!htb]
\centering
  \pgfplotslegendfromname{legend:Cartesian-vs-curved-PID-Euler}
  \\
  \begin{subfigure}{0.49\linewidth}
    \begin{tikzpicture}[
      font=\footnotesize
      ]
      \begin{axis}[
          xlabel={Polynomial degree},
          ylabel={Time/RHS/DOF {[sec]}},
          width=\textwidth,
          height=0.66\textwidth,
          ymin=0.0,
          grid=major,
        ]
        \addplot+ [error bars/.cd, y dir=both,y explicit,]
          table [x index=0, y index=1, y error index=2]{code/Cartesian_vs_curved/pids_2D_flux_shima_etal.dat};

        \addplot+ [error bars/.cd, y dir=both,y explicit,]
          table [x index=0, y index=3, y error index=4]{code/Cartesian_vs_curved/pids_2D_flux_shima_etal.dat};

        \addplot+ [error bars/.cd, y dir=both,y explicit,]
          table [x index=0, y index=5, y error index=6]{code/Cartesian_vs_curved/pids_2D_flux_shima_etal.dat};
      \end{axis}
    \end{tikzpicture}%
    \caption{2D, flux of Shima \etal \eqref{eq:flux_shima_etal}.}
  \end{subfigure}%
  \hspace*{\fill}
  \begin{subfigure}{0.49\linewidth}
    \begin{tikzpicture}[
      font=\footnotesize
      ]
      \begin{axis}[
          xlabel={Polynomial degree},
          ylabel={Time/RHS/DOF {[sec]}},
          width=\textwidth,
          height=0.66\textwidth,
          ymin=0.0,
          grid=major,
        ]
        \addplot+ [error bars/.cd, y dir=both,y explicit,]
          table [x index=0, y index=1, y error index=2]{code/Cartesian_vs_curved/pids_3D_flux_shima_etal.dat};

        \addplot+ [error bars/.cd, y dir=both,y explicit,]
          table [x index=0, y index=3, y error index=4]{code/Cartesian_vs_curved/pids_3D_flux_shima_etal.dat};

        \addplot+ [error bars/.cd, y dir=both,y explicit,]
          table [x index=0, y index=5, y error index=6]{code/Cartesian_vs_curved/pids_3D_flux_shima_etal.dat};

        \addplot+ [error bars/.cd, y dir=both,y explicit]
          table [x index=0, y index=1, y error index=2, col sep=comma]{code/ec_performance_fluxo/serialTests/results/results_Shima.dat};
      \end{axis}
    \end{tikzpicture}%
    \caption{3D, flux of Shima \etal \eqref{eq:flux_shima_etal}.}
  \end{subfigure}%
  \\
  \begin{subfigure}{0.49\linewidth}
    \begin{tikzpicture}[
      font=\footnotesize
      ]
      \begin{axis}[
          xlabel={Polynomial degree},
          ylabel={Time/RHS/DOF {[sec]}},
          width=\textwidth,
          height=0.66\textwidth,
          ymin=0.0,
          grid=major,
        ]
        \addplot+ [error bars/.cd, y dir=both,y explicit,]
          table [x index=0, y index=1, y error index=2]{code/Cartesian_vs_curved/pids_2D_flux_ranocha.dat};

        \addplot+ [error bars/.cd, y dir=both,y explicit,]
          table [x index=0, y index=3, y error index=4]{code/Cartesian_vs_curved/pids_2D_flux_ranocha.dat};

        \addplot+ [error bars/.cd, y dir=both,y explicit,]
          table [x index=0, y index=5, y error index=6]{code/Cartesian_vs_curved/pids_2D_flux_ranocha.dat};
      \end{axis}
    \end{tikzpicture}%
    \caption{2D, flux of Ranocha \eqref{eq:flux_ranocha}.}
  \end{subfigure}%
  \hspace*{\fill}
  \begin{subfigure}{0.49\linewidth}
    \begin{tikzpicture}[
      font=\footnotesize
      ]
      \begin{axis}[
          xlabel={Polynomial degree},
          ylabel={Time/RHS/DOF {[sec]}},
          width=\textwidth,
          height=0.66\textwidth,
          legend to name=legend:Cartesian-vs-curved-PID-Euler,
          legend columns=-1,
          ymin=0.0,
          grid=major,
        ]
        \addplot+ [error bars/.cd, y dir=both,y explicit,]
          table [x index=0, y index=1, y error index=2]{code/Cartesian_vs_curved/pids_3D_flux_ranocha.dat};
          \addlegendentry{\trixi, \codeinline{TreeMesh}}

        \addplot+ [error bars/.cd, y dir=both,y explicit,]
          table [x index=0, y index=3, y error index=4]{code/Cartesian_vs_curved/pids_3D_flux_ranocha.dat};
          \addlegendentry{\trixi, \codeinline{StructuredMesh}}

        \addplot+ [error bars/.cd, y dir=both,y explicit,]
          table [x index=0, y index=5, y error index=6]{code/Cartesian_vs_curved/pids_3D_flux_ranocha.dat};
          \addlegendentry{\trixi, \codeinline{P4estMesh}}

        \addplot+ [error bars/.cd, y dir=both,y explicit]
          table [x index=0, y index=1, y error index=2, col sep=comma]{code/ec_performance_fluxo/serialTests/results/results_Ranocha.dat};
          \addlegendentry{\fluxo}
      \end{axis}
    \end{tikzpicture}%
    \caption{3D, flux of Ranocha \eqref{eq:flux_ranocha}.}
  \end{subfigure}%
  \caption{Runtime per right-hand side evaluation and degree of freedom for
           different mesh types and fluxes, using the LGL-DGSEM discretization
           with flux differencing of the compressible Euler equations.}
  \label{fig:Cartesian-vs-curved-PID-Euler}
\end{figure}

The baseline \PID results are visualized in
Figure~\ref{fig:Cartesian-vs-curved-PID-Euler}.
As expected based on the number of operations, the PID increases linearly
with the polynomial degree $\polydeg$ (see Table~\ref{tab:computational-complexity};
the computational complexity $\O\bigl( \ndims (\polydeg + 1)^{\ndims + 1} \bigr)$
needs to be divided by the number of DOFs per element, $(\polydeg + 1)^{\ndims}$,
to get the scaling of the \PID).
The current handling of unstructured meshes in the \codeinline{P4estMesh} of
\trixi compared to the \codeinline{StructuredMesh} has no visible impact in 2D
and only a minor impact of approx.\ \SI{5}{\percent} in 3D. In contrast,
the Cartesian \codeinline{TreeMesh} can improve the performance significantly
by up to \SI{33}{\percent} for cheap volume fluxes and
up to \SI{20}{\percent} for expensive volume fluxes involving logarithmic mean values;
the impact in 2D is reduced by approx.\ ten percentage points.

Moreover, these benchmarks show that the serial performance results of \fluxo
(written in Fortran) and \trixi (written in Julia) are similar.
These comparisons are based on compiling \fluxo with all available performance
tuning options, including explicit inlining of the volume fluxes and setting
the polynomial degree (and node type) as constant at compile time.
Thus, these numerical results demonstrate that Julia can be used for
performance-critical scientific computing. Therefore, we will present most
microbenchmarks using only Julia code in the following. This also simplifies
the presentation in the accompanying repository \cite{ranocha2021efficientRepro}
since it is easier to work with a high-level language (Julia) and a library-based
approach (\trixi).

\subsection{Different versions of numerical fluxes}
\label{sec:numflux-benchmarks}

We perform microbenchmarks comparing different versions of numerical fluxes
for the compressible Euler equations. In particular, we benchmark the optimized
directional fluxes presented in Section~\ref{sec:numflux-Euler} and their
corresponding Cartesian versions. Since the compressible Euler equations are rotationally invariant,
a common approach to compute numerical fluxes in arbitrary directions is to rotate
the states into the first coordinate direction, compute the standard Cartesian flux
there, and rotate the resulting flux back, see \cite[Section~16.7.3]{toro2009riemann}.
We also benchmark this approach including the on-the-fly computation of an
appropriate rotation matrix (tangent vectors) from a given normal direction/vector,
see Algorithm~\ref{alg:rotated-flux}.
Additionally, we benchmark an optimized alternative thereof with precomputed
rotation matrix.

\begin{algorithm}[!hbt]
  \caption{Computation of rotated numerical fluxes.}
  \label{alg:rotated-flux}
  \begin{algorithmic}
    \Require Input states $u_{\mathrm{in}}, u_{\mathrm{out}}$, normal direction $\tilde\normal$, Cartesian numerical flux $\fnum$
    \Ensure Numerical flux $\fnum(u_{\mathrm{in}}, u_{\mathrm{out}}; \tilde\normal)$
    \State Normalize the normal direction $\normal = \tilde\normal / \| \tilde\normal \|$
    \State Compute $\ndims - 1$ orthonormal tangent vectors $t_i \perp \normal$ to get the rotation matrix $(\normal, t_i)^T$
    \State Rotate states $u_{\mathrm{in}}, u_{\mathrm{out}}$ to first coordinate direction, resulting in rotated states $u_-, u_+$
    \State Compute the Cartesian numerical flux $\fnum{1}(u_-, u_+)$
    \State Rotate the flux $\fnum{1}(u_-, u_+)$ back from first coordinate direction to obtain $\fnum(u_{\mathrm{in}}, u_{\mathrm{out}}; \normal)$
    \State \Return $\fnum(u_{\mathrm{in}}, u_{\mathrm{out}}; \tilde\normal) = \fnum(u_{\mathrm{in}}, u_{\mathrm{out}}; \normal) \| \tilde\normal \|$
  \end{algorithmic}
\end{algorithm}

\begin{table}[htb]
\centering
\caption{Microbenchmarks of different versions of numerical fluxes for the
         compressible Euler equations.}
\label{tab:numflux-benchmarks}
\begin{tabular*}{\linewidth}{@{\extracolsep{\fill}} l *4c @{}}
  \toprule
    & Cartesian
    & Directional
    & \multicolumn{2}{c}{Rotated}
  \\
    &
    &
    & (on the fly)
    & (precomputed)
  \\
  \midrule
  Flux of Shima \etal \eqref{eq:flux_shima_etal}, 2D
    & \SI{ 7.7 \pm 0.1}{\ns}  
    & \SI{ 9.1 \pm 0.2}{\ns}  
    & \SI{18.0 \pm 0.3}{\ns}  
    & \SI{11.8 \pm 0.2}{\ns}  
  \\
  Flux of Shima \etal \eqref{eq:flux_shima_etal}, 3D
    & \SI{ 9.5 \pm 0.2}{\ns}  
    & \SI{12.1 \pm 0.2}{\ns}  
    & \SI{53.4 \pm 0.4}{\ns}  
    & \SI{18.7 \pm 0.3}{\ns}  
  \\
  Flux of Ranocha \eqref{eq:flux_ranocha}, 2D
    & \SI{33.2 \pm 0.3}{\ns}  
    & \SI{34.8 \pm 0.3}{\ns}  
    & \SI{42.1 \pm 0.3}{\ns}  
    & \SI{37.4 \pm 0.3}{\ns}  
  \\
  Flux of Ranocha \eqref{eq:flux_ranocha}, 3D
    & \SI{35.4 \pm 0.3}{\ns}  
    & \SI{39.7 \pm 0.3}{\ns}  
    & \SI{82.1 \pm 0.4}{\ns}  
    & \SI{46.0 \pm 0.3}{\ns}  
  \\
  LLF flux, 2D
    & \SI{18.3 \pm 0.3}{\ns}  
    & \SI{19.9 \pm 0.2}{\ns}  
    & \SI{28.3 \pm 0.2}{\ns}  
    & \SI{21.2 \pm 0.2}{\ns}  
  \\
  LLF flux, 3D
    & \SI{19.7 \pm 0.2}{\ns}  
    & \SI{21.5 \pm 0.2}{\ns}  
    & \SI{68.2 \pm 0.4}{\ns}  
    & \SI{27.7 \pm 0.4}{\ns}  
  \\
  HLL flux, 2D
    & \SI{21.1 \pm 0.2}{\ns}  
    & \SI{21.4 \pm 0.2}{\ns}  
    & \SI{31.6 \pm 0.3}{\ns}  
    & \SI{25.4 \pm 0.3}{\ns}  
  \\
  HLL flux, 3D
    & \SI{23.3 \pm 0.2}{\ns}  
    & \SI{22.9 \pm 0.3}{\ns}  
    & \SI{79.0 \pm 0.4}{\ns}  
    & \SI{34.6 \pm 0.3}{\ns}  
  \\
  \bottomrule
\end{tabular*}
\end{table}

The results of these microbenchmarks are shown in Table~\ref{tab:numflux-benchmarks}.
The symmetric numerical fluxes used for the volume terms behave as follows.
The Cartesian fluxes are usually the most efficient versions, followed directly
by the directional approach without rotation. Depending on the computational
complexity of the numerical flux, the Cartesian version is between
\SI{5}{\percent} and \SI{20}{\percent}
more efficient than the directional version.
The rotated version with precomputed terms is significantly more expensive than
the directional version, usually between
\SI{10}{\percent} (2D, expensive flux) and \SI{2}{\times} (3D, cheap flux).
Finally, the on-the-fly version without precomputed rotation matrix is significantly
more expensive than the rotated version with precomputed terms,
approximately between
\SI{10}{\percent} (2D, expensive flux) and \SI{3}{\times} (3D, cheap flux).

For these comparisons we also benchmarked simple versions of the local Lax-Friedrichs/Rusanov flux (LLF)
and the HLL flux \cite{harten1983upstream}. Such numerical fluxes are usually
used at interfaces to introduce additional dissipation. The runtimes of these
fluxes is between the cheap flux of Shima \etal \eqref{eq:flux_shima_etal} and
the expensive EC flux of Ranocha \eqref{eq:flux_ranocha}. Thus,
we continue to focus on the volume terms using flux differencing in this article.

The key messages of these microbenchmarks are as follows. First, the improved
performance of the Cartesian mesh reported in Section~\ref{sec:Cartesian-vs-curved}
is not only caused by the different versions of numerical fluxes but also by
the reduced amount of operations necessary to deal with general curvilinear coordinates.
Second, a direct implementation of directional numerical fluxes
is often preferable compared to an implementation that uses rotations; if the
latter should nevertheless be used, the necessary rotation terms should be
computed in advance, particularly for 3D implementations.

\section{Comparison to overintegration}
\label{sec:overintegration}

Another common strategy to increase the robustness of LGL-DGSEM discretizations is
overintegration, \ie, interpolating the numerical solution to a higher polynomial
degree, using standard weak form volume terms there, and projecting orthogonally
on the given polynomial degree. A comparison of overintegration and flux
differencing based DG methods for under-resolved turbulence is presented in
\cite{winters2018comparative}.

Here, we perform microbenchmarks of the volume terms using flux differencing LGL-DGSEM
and overintegration with different polynomial degrees. For the overintegration,
we follow the procedure presented in, e.g., \cite{gassner2013accuracy}. The interpolation and
projection steps use sum factorization and efficient multiplication kernels
using tools from LoopVectorization.jl\footnote{\url{https://github.com/JuliaSIMD/LoopVectorization.jl}},
which is on par with (and sometimes faster than) optimized BLAS libraries such
as Intel MKL for matrix multiplications at these sizes \cite{elrod2021roadmap}.
The numerical solution is initialized on a Cartesian \codeinline{TreeMesh} with
a single element for the isentropic vortex initial condition described in
Section~\ref{sec:Cartesian-vs-curved}.

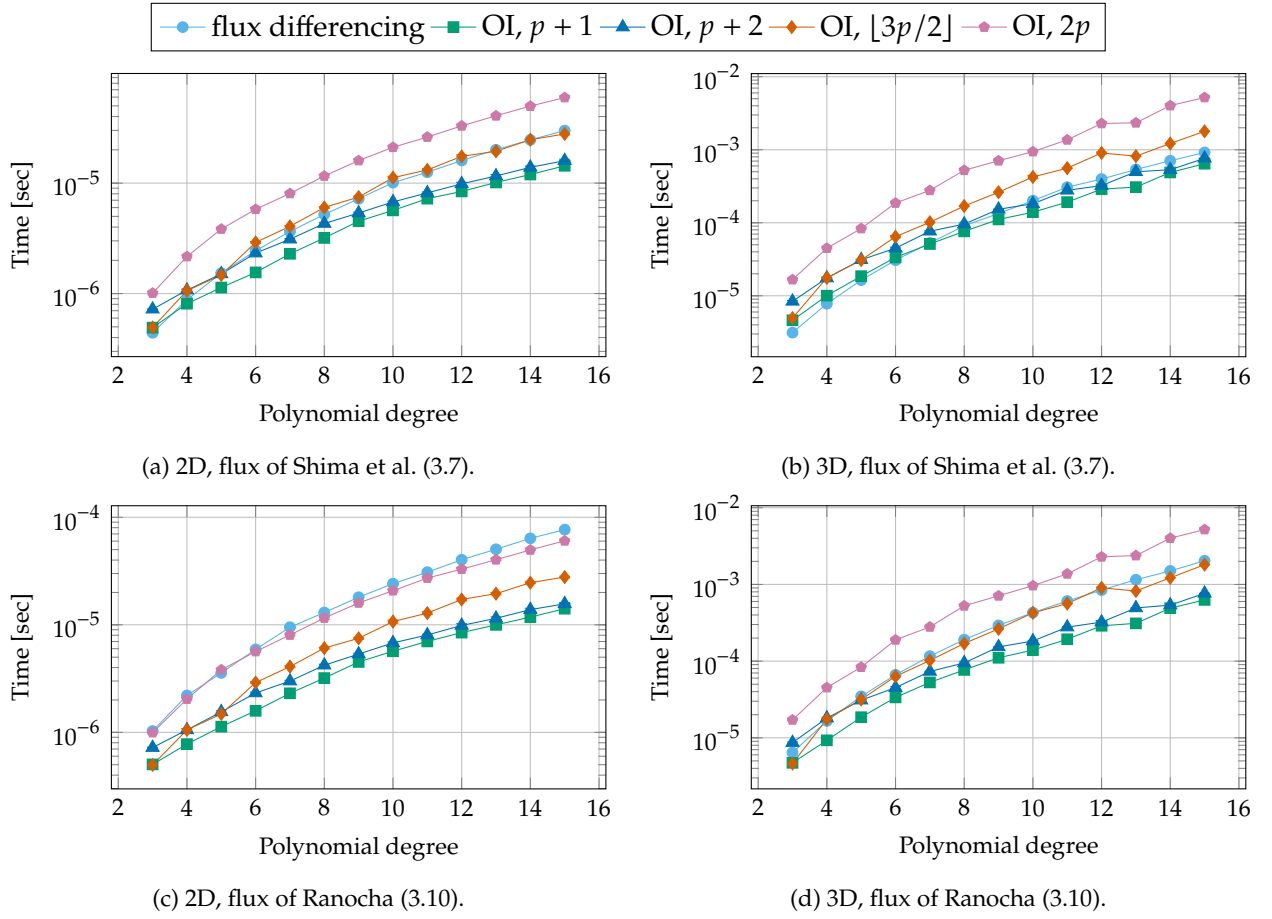
\begin{figure}[!htb]
\centering
  \pgfplotslegendfromname{legend:overintegration-Euler}
  \\
  \begin{subfigure}{0.49\linewidth}
    \begin{tikzpicture}[
      font=\footnotesize
      ]
      \begin{semilogyaxis}[
          xlabel={Polynomial degree},
          ylabel={Time {[sec]}},
          width=\textwidth,
          height=0.66\textwidth,
          legend columns=-1,
          legend to name=legend:overintegration-Euler,
          grid=major,
        ]
        \addplot+ [error bars/.cd, y dir=both,y explicit,]
          table [x index=0, y index=1, y error index=2]{code/overintegration/overintegration_2D_flux_shima_etal.dat};
          \addlegendentry{flux differencing}

        \addplot+ [error bars/.cd, y dir=both,y explicit,]
          table [x index=0, y index=3, y error index=4]{code/overintegration/overintegration_2D_flux_shima_etal.dat};
          \addlegendentry{OI, $\polydeg + 1$}

        \addplot+ [error bars/.cd, y dir=both,y explicit,]
          table [x index=0, y index=5, y error index=6]{code/overintegration/overintegration_2D_flux_shima_etal.dat};
          \addlegendentry{OI, $\polydeg + 2$}

        \addplot+ [error bars/.cd, y dir=both,y explicit,]
          table [x index=0, y index=7, y error index=8]{code/overintegration/overintegration_2D_flux_shima_etal.dat};
          \addlegendentry{OI, $\lfloor 3 \polydeg / 2 \rfloor$}

        \addplot+ [error bars/.cd, y dir=both,y explicit,]
          table [x index=0, y index=9, y error index=10]{code/overintegration/overintegration_2D_flux_shima_etal.dat};
          \addlegendentry{OI, $2 \polydeg$}
      \end{semilogyaxis}
    \end{tikzpicture}%
    \caption{2D, flux of Shima \etal \eqref{eq:flux_shima_etal}.}
  \end{subfigure}%
  \hspace*{\fill}
  \begin{subfigure}{0.49\linewidth}
    \begin{tikzpicture}[
      font=\footnotesize
      ]
      \begin{semilogyaxis}[
          xlabel={Polynomial degree},
          ylabel={Time {[sec]}},
          width=\textwidth,
          height=0.66\textwidth,
          grid=major,
        ]
        \addplot+ [error bars/.cd, y dir=both,y explicit,]
          table [x index=0, y index=1, y error index=2]{code/overintegration/overintegration_3D_flux_shima_etal.dat};

        \addplot+ [error bars/.cd, y dir=both,y explicit,]
          table [x index=0, y index=3, y error index=4]{code/overintegration/overintegration_3D_flux_shima_etal.dat};

        \addplot+ [error bars/.cd, y dir=both,y explicit,]
          table [x index=0, y index=5, y error index=6]{code/overintegration/overintegration_3D_flux_shima_etal.dat};

        \addplot+ [error bars/.cd, y dir=both,y explicit,]
          table [x index=0, y index=7, y error index=8]{code/overintegration/overintegration_3D_flux_shima_etal.dat};

        \addplot+ [error bars/.cd, y dir=both,y explicit,]
          table [x index=0, y index=9, y error index=10]{code/overintegration/overintegration_3D_flux_shima_etal.dat};
      \end{semilogyaxis}
    \end{tikzpicture}%
    \caption{3D, flux of Shima \etal \eqref{eq:flux_shima_etal}.}
  \end{subfigure}%
  \\
  \begin{subfigure}{0.49\linewidth}
    \begin{tikzpicture}[
      font=\footnotesize
      ]
      \begin{semilogyaxis}[
          xlabel={Polynomial degree},
          ylabel={Time {[sec]}},
          width=\textwidth,
          height=0.66\textwidth,
          grid=major,
        ]
        \addplot+ [error bars/.cd, y dir=both,y explicit,]
          table [x index=0, y index=1, y error index=2]{code/overintegration/overintegration_2D_flux_ranocha.dat};

        \addplot+ [error bars/.cd, y dir=both,y explicit,]
          table [x index=0, y index=3, y error index=4]{code/overintegration/overintegration_2D_flux_ranocha.dat};

        \addplot+ [error bars/.cd, y dir=both,y explicit,]
          table [x index=0, y index=5, y error index=6]{code/overintegration/overintegration_2D_flux_ranocha.dat};

        \addplot+ [error bars/.cd, y dir=both,y explicit,]
          table [x index=0, y index=7, y error index=8]{code/overintegration/overintegration_2D_flux_ranocha.dat};

        \addplot+ [error bars/.cd, y dir=both,y explicit,]
          table [x index=0, y index=9, y error index=10]{code/overintegration/overintegration_2D_flux_ranocha.dat};
      \end{semilogyaxis}
    \end{tikzpicture}%
    \caption{2D, flux of Ranocha \eqref{eq:flux_ranocha}.}
  \end{subfigure}%
  \hspace*{\fill}
  \begin{subfigure}{0.49\linewidth}
    \begin{tikzpicture}[
      font=\footnotesize
      ]
      \begin{semilogyaxis}[
          xlabel={Polynomial degree},
          ylabel={Time {[sec]}},
          width=\textwidth,
          height=0.66\textwidth,
          grid=major,
        ]
        \addplot+ [error bars/.cd, y dir=both,y explicit,]
          table [x index=0, y index=1, y error index=2]{code/overintegration/overintegration_3D_flux_ranocha.dat};

        \addplot+ [error bars/.cd, y dir=both,y explicit,]
          table [x index=0, y index=3, y error index=4]{code/overintegration/overintegration_3D_flux_ranocha.dat};

        \addplot+ [error bars/.cd, y dir=both,y explicit,]
          table [x index=0, y index=5, y error index=6]{code/overintegration/overintegration_3D_flux_ranocha.dat};

        \addplot+ [error bars/.cd, y dir=both,y explicit,]
          table [x index=0, y index=7, y error index=8]{code/overintegration/overintegration_3D_flux_ranocha.dat};

        \addplot+ [error bars/.cd, y dir=both,y explicit,]
          table [x index=0, y index=9, y error index=10]{code/overintegration/overintegration_3D_flux_ranocha.dat};
      \end{semilogyaxis}
    \end{tikzpicture}%
    \caption{3D, flux of Ranocha \eqref{eq:flux_ranocha}.}
  \end{subfigure}%
  \caption{Microbenchmarks of overintegration (OI) vs. flux differencing
           volume terms of DG discretizations with polynomials of degree $\polydeg$
           for the  the compressible Euler equations.}
  \label{fig:overintegration-Euler}
\end{figure}

The results are visualized in Figure~\ref{fig:overintegration-Euler}.
The three-dimensional case is most relevant in practice. There, flux
differencing with an inexpensive volume flux, such as the one of Shima \etal
\eqref{eq:flux_shima_etal}, is cheaper than any overintegration for small polynomial
degrees $3 \le \polydeg \le 5$. For $5 < \polydeg \le 7$, flux differencing is between
overintegration with one or two additional nodes per coordinate direction.
For an expensive volume flux involving logarithmic mean values, flux differencing
is still between overintegration with one or two additional nodes per coordinate
direction for $\polydeg \in \{3, 4\}$. For $\polydeg > 5$, flux differencing
is between overintegration with internal polynomial degrees of
$\lfloor 3 \polydeg / 2 \rfloor$ and $2 \polydeg$.
In 2D, flux differencing is relatively more expensive than in 3D. Nevertheless,
cheap volume fluxes make it still faster than overintegration with $2 \polydeg$.

The key message of these benchmarks is that flux differencing is competetive
with overintegration, in particular in 3D, and it has stability guarantees that
the overintegration strategy typically does not have.
For the most practically relevant case
for computational fluid dynamics (3D, polynomial degree $\le 3$), flux differencing
can be faster than overintegration of the volume terms with a single additional
degree of freedom per coordinate direction.
Moreover, there exist several cases where
overintegration fails to provide appropriate robustness while flux differencing
schemes are stable \cite{winters2018comparative}. Additionally, analogous to the
comparison of Legendre-Gauss and Legendre-Gauss-Lobatto quadrature rules in discontinuous Galerkin spectral
element methods, overintegration comes at the cost of increased stiffness of the
semidiscretization, resulting roughly in a factor of two in time step restrictions
of explicit time integration methods \cite{gassner2011comparison}.
Moreover, many overintegration variants apply a similar procedure also to surface
terms, making them more expensive than the standard surface terms used for flux
differencing methods.

\section{Gauss collocation methods and entropy projections}
\label{sec:Gauss}

In addition to entropy stable schemes based on (\ref{eq:fluxdiff-SBP}), it is possible
to construct entropy stable schemes based on generalized SBP operators
\cite{fernandez2014generalized}. These include, for example, collocation schemes constructed
on Legendre-Gauss nodes \cite{chan2019efficient}. These schemes can be written in the
form \cite{chan2019skew}
\begin{equation}
\label{eq:fluxdiff-GSBP-opDsplit}
  \partial_t \vec{u}
  + \widetilde{\VOL}
  + \opM^{-1} \opR^T \opB \vecfnum
  = \vec{0}.
\end{equation}
The operators $\opM, \opR, \opD{j}$ are defined as in Section~\ref{sec:SBP}. However, one
key difference between \eqref{eq:fluxdiff-SBP} and \eqref{eq:fluxdiff-GSBP-opDsplit}
is that it is no longer assumed that volume nodes include nodes on the boundary. Instead, the
boundary restriction operator $\opR$ now maps from interior nodes to boundary nodes,
resulting in a fully dense matrix. As a result, the volume terms must be modified to retain both
high order accuracy and entropy stability. Recall that ${u}$ denotes the mapping from entropy
variables to conservative variables and $w$ denotes the mapping from conservative variables
to entropy variables. Then, the volume terms $\widetilde{\VOL}$ are computed via
\begin{equation}
\begin{gathered}
\widetilde{\VOL} = \opM^{-1}\begin{bmatrix}
\I \\
\opR
\end{bmatrix}^T
\vec{f}^{\mathrm{hybrid}}, \qquad
\vec{f}^{\mathrm{hybrid}}_i = \sum_{j=1}^\ndims \sum_k 2 (\opQ{h,j})_{i,k} \fvol{j}(\vec{\tilde{u}}_i, \vec{\tilde{u}}_k), \\
\opQ{h,j} = \frac{1}{2}\begin{bmatrix}
\opM \opD{j} - (\opM\opD{j})^T & \opR^T B N_j\\
-B N_j \opR & 0
\end{bmatrix},
\qquad
\vec{\tilde{u}} = \begin{bmatrix}
\vec{u}\\
{u}\bigl(\opR w(\vec{u})\bigr)
\end{bmatrix},
\end{gathered}
\label{eq:hybridized_gauss_formulation}
\end{equation}
where $k$ sums over the combined set of both volume and surface quadrature points.
Here, $\vec{\tilde{u}}$ is the ``entropy projection'', and $\opQ{h,j}$ denotes the
\textit{hybridized} SBP operator with respect to the $j$th
coordinate direction \cite{chan2018discretely}.
Entropy stable schemes constructed on the Legendre-Gauss nodes can be alternatively
formulated in terms of ``correction'' terms on the surface, allowing an implementation
which keeps the volume term $\VOL$ from \eqref{eq:fluxdiff-SBP} untouched:
\begin{equation}
\label{eq:fluxdiff-GSBP-surfCorr}
\begin{gathered}
  \partial_t \vec{u}
  + \VOL
  + \opM^{-1} \left(\opR^T \opB \vecfnum +
  \begin{bmatrix}
  \I \\
  \opR
  \end{bmatrix}^T
 \vecfcorr \right)
  = \vec{0},\\
  \vecfcorr_i = \sum_{j=1}^\ndims \sum_k (\opB{h,j})_{i,k} \fvol{j}(\vec{\tilde{u}}_i, \vec{\tilde{u}}_k),
 \qquad
  \opB{h,j} =
\begin{bmatrix}
   0 & \opR^T B N_j\\
   -B N_j \opR & 0
\end{bmatrix},
\end{gathered}
\end{equation}
where $k$ sums again over the combined set of both volume and surface quadrature
points and $\opB{h,j}$ denotes a \textit{hybridized} boundary matrix.

The efficient implementation of \eqref{eq:fluxdiff-GSBP-opDsplit} and \eqref{eq:fluxdiff-GSBP-surfCorr}
needs special care to avoid unnecessary flux evaluations and multiplications by zero.
This can be achieved by using sparse matrix formats and the tensor product structure
within both Legendre-Gauss and Legendre-Gauss-Lobatto bases on tensor product elements.
We note that \trixi uses \eqref{eq:fluxdiff-GSBP-opDsplit} for the implementation, while
FLUXO uses \eqref{eq:fluxdiff-GSBP-surfCorr}.
In the results of this section, we only use the entropy-conservative scheme,
using the flux of Ranocha~\eqref{eq:flux_ranocha} in the flux differencing volume
terms and for the surface fluxes.

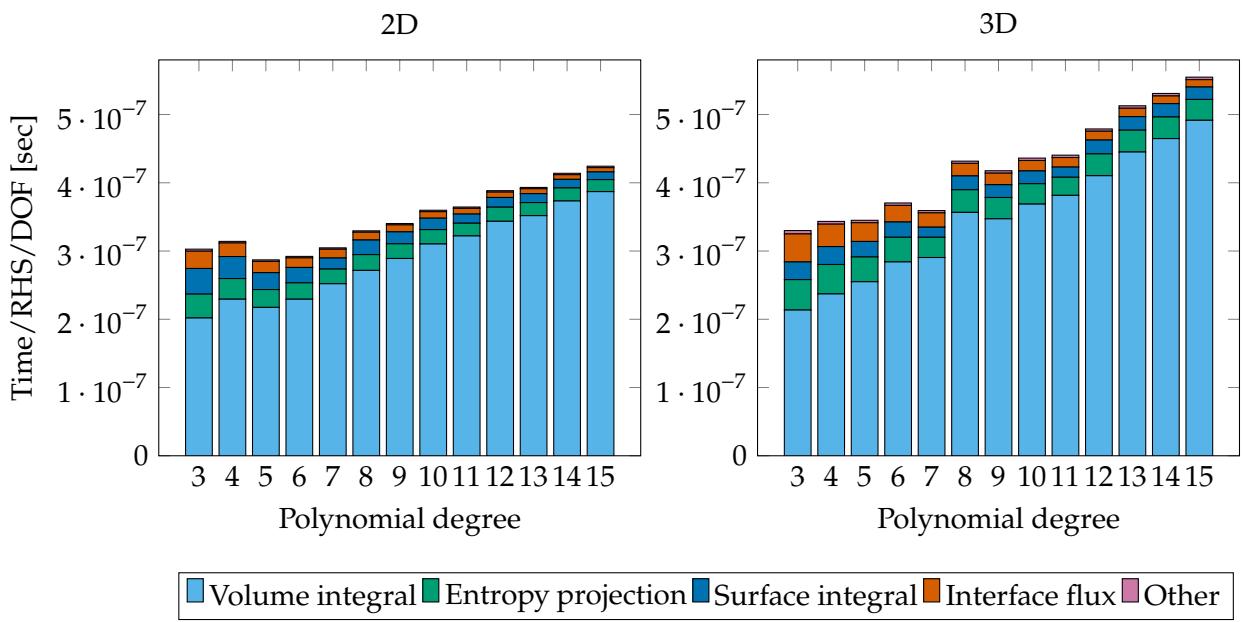
\begin{figure}[!htb]
\centering
    \begin{tikzpicture}
    \begin{groupplot}[
      group style={
      group size=2 by 1,
      horizontal sep=4em,
      group name=plots
    },
      width=.48\textwidth,
      ybar stacked,
      xtick=data,
      scaled y ticks=false,
      xlabel={Polynomial degree},
      ylabel={Time/RHS/DOF {[sec]}},
      legend columns=-1,
      ymin={0},
      ymax={5.8e-7},
    ]
      \nextgroupplot[legend to name=legend:GaussPID, title={2D}]
      \addplot [fill=mycolor1] table [y=time_volume, x=Polydeg] {code/Gauss/Gauss_2D_relative_timings.dat};
      \addlegendentry{Volume integral}
      \addplot [fill=mycolor2] table [y=time_entropy_projection, x=Polydeg] {code/Gauss/Gauss_2D_relative_timings.dat};
      \addlegendentry{Entropy projection}
      \addplot [fill=mycolor3] table [y=time_surface, x=Polydeg] {code/Gauss/Gauss_2D_relative_timings.dat};
      \addlegendentry{Surface integral}
      \addplot [fill=mycolor4] table [y=time_interface, x=Polydeg] {code/Gauss/Gauss_2D_relative_timings.dat};
      \addlegendentry{Interface flux}
      \addplot [fill=mycolor5] table [y=other, x=Polydeg] {code/Gauss/Gauss_2D_relative_timings.dat};
      \addlegendentry{Other}

      \nextgroupplot[ylabel=\empty, title={3D}]
      \addplot [fill=mycolor1] table [y=time_volume, x=Polydeg] {code/Gauss/Gauss_3D_relative_timings.dat};
      \addplot [fill=mycolor2] table [y=time_entropy_projection, x=Polydeg] {code/Gauss/Gauss_3D_relative_timings.dat};
      \addplot [fill=mycolor3] table [y=time_surface, x=Polydeg] {code/Gauss/Gauss_3D_relative_timings.dat};
      \addplot [fill=mycolor4] table [y=time_interface, x=Polydeg] {code/Gauss/Gauss_3D_relative_timings.dat};
      \addplot [fill=mycolor5] table [y=other, x=Polydeg] {code/Gauss/Gauss_3D_relative_timings.dat};
    \end{groupplot}

    \node at (plots c2r1.south) [inner sep=0pt, anchor=north, xshift=-10em, yshift=-4em] {\pgfplotslegendfromname{legend:GaussPID}};

    \end{tikzpicture}
    \caption{\PID breakdowns for two and three dimensional benchmarks using a Legendre-Gauss collocation solver.}
    \label{fig:GaussPID}
\end{figure}

Figure~\ref{fig:GaussPID} shows \PID values (broken down by subroutines)
for 2D and 3D Legendre-Gauss collocation solvers of degree $3, \ldots, 15$ in Trixi.jl.
For all polynomial degrees studied here, we observe that the volume integral
strongly dominates computational runtime. This step is dominated by flux evaluations,
but also includes a ``lifting'' of face contributions to volume nodes when computing
(\ref{eq:hybridized_gauss_formulation}) or (\ref{eq:fluxdiff-GSBP-surfCorr}).
The next most expensive step is the entropy
projection, which makes up a smaller share of the overall runtime as $\polydeg$ increases.
For example, for $\polydeg > 10$ in both two and three dimensions, the cost of the entropy
projection is close to the combined cost of all subroutines aside from flux differencing volume terms.

The relative costs (as percentage of \PID) of flux
differencing and the entropy projection depend strongly on the polynomial degree
$\polydeg$; a higher polynomial degree increases the relative cost of flux differencing
compared to the entropy projection.
In 2D, these relative costs range from
approx.\ \SI{66}{\percent} (flux differencing) and \SI{11}{\percent} (entropy projection)
for $\polydeg = 3$ to
approx.\ \SI{91}{\percent} (flux differencing) and \SI{4}{\percent} (entropy projection)
for $\polydeg = 15$.
In 3D, these relative costs range from
approx.\ \SI{65}{\percent} (flux differencing) and \SI{14}{\percent} (entropy projection)
for $\polydeg = 3$ to
approx.\ \SI{89}{\percent} (flux differencing) and \SI{6}{\percent} (entropy projection)
for $\polydeg = 15$.

\begin{figure}[!htb]
\centering
  \pgfplotslegendfromname{legend:Gauss}
  \\
  \begin{subfigure}{0.49\linewidth}
    \begin{tikzpicture}[
      font=\footnotesize
      ]
      \begin{axis}[
          xlabel={Polynomial degree},
          ylabel={Time/RHS/DOF {[sec]}},
          width=\textwidth,
          height=0.66\textwidth,
          ymin=0.0,
          grid=major,
        ]
        \addplot+ [error bars/.cd, y dir=both,y explicit,]
          table [x index=0, y index=1, y error index=2]{code/Gauss/Gauss_2D_flux_ranocha.dat};

        \addplot+ [error bars/.cd, y dir=both,y explicit,]
          table [x index=0, y index=3, y error index=4]{code/Gauss/Gauss_2D_flux_ranocha.dat};

        \addplot+ [error bars/.cd, y dir=both,y explicit,]
          table [x index=0, y index=5, y error index=6]{code/Cartesian_vs_curved/pids_2D_flux_ranocha.dat};
      \end{axis}
    \end{tikzpicture}%
    \caption{2D.}
  \end{subfigure}%
  \hspace*{\fill}
  \begin{subfigure}{0.49\linewidth}
    \begin{tikzpicture}[
      font=\footnotesize
      ]
      \begin{axis}[
          xlabel={Polynomial degree},
          ylabel={Time/RHS/DOF {[sec]}},
          width=\textwidth,
          height=0.66\textwidth,
          legend columns=-1,
          legend to name=legend:Gauss,
          ymin=0.0,
          grid=major,
        ]
        \addplot+ [error bars/.cd, y dir=both,y explicit,]
          table [x index=0, y index=1, y error index=2]{code/Gauss/Gauss_3D_flux_ranocha.dat};
          \addlegendentry{Trixi.jl, \codeinline{SBP}}

        \addplot+ [error bars/.cd, y dir=both,y explicit,]
          table [x index=0, y index=3, y error index=4]{code/Gauss/Gauss_3D_flux_ranocha.dat};
          \addlegendentry{Trixi.jl, \codeinline{GaussSBP}}

        \addplot+ [error bars/.cd, y dir=both,y explicit,]
          table [x index=0, y index=5, y error index=6]{code/Cartesian_vs_curved/pids_3D_flux_ranocha.dat};
          \addlegendentry{Trixi.jl, \codeinline{P4estMesh}}

        \addplot+ [error bars/.cd, y dir=both,y explicit]
          table [x index=0, y index=1, y error index=2, col sep=comma]{code/ec_performance_fluxo/serialTests/results/results_Ranocha_Gauss.dat};
          \addlegendentry{FLUXO, \codeinline{GaussSBP}}
      \end{axis}
    \end{tikzpicture}%
    \caption{3D.}
  \end{subfigure}%
  \caption{Runtime per right-hand side evaluation and degree of freedom for
           different mesh/node types and entropy-conservative DG discretizations
           using the flux \eqref{eq:flux_ranocha} for the compressible Euler
           equations.}
  \label{fig:Gauss}
\end{figure}
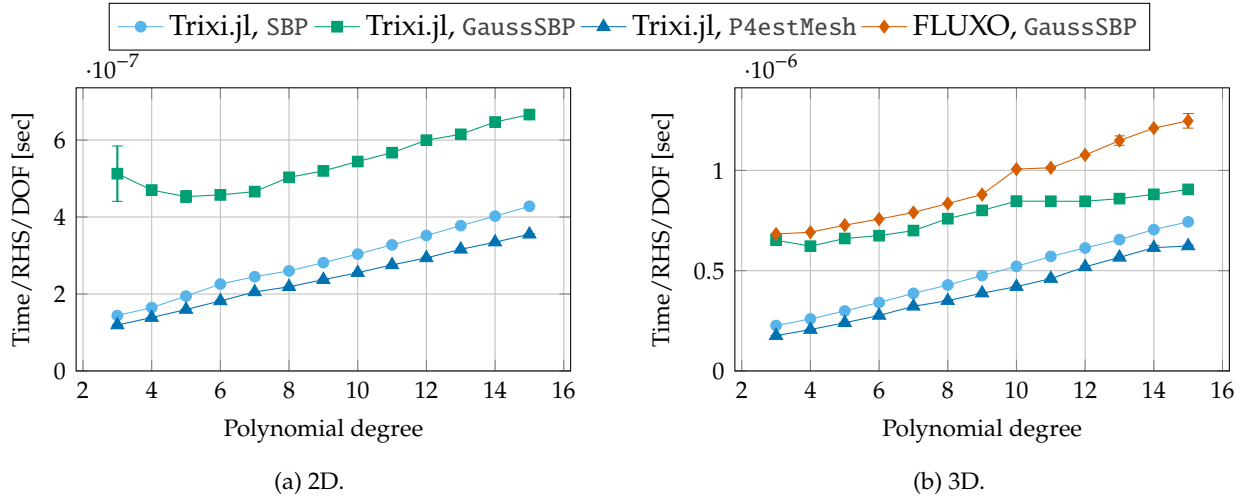

Figure~\ref{fig:Gauss} compares
\PID results for several different implementations of entropy-conservative LGL-DGSEM
(Trixi.jl, \codeinline{SBP}, which uses affine mappings, and \codeinline{P4estMesh}, which uses curvilinear mappings)
and Legendre-Gauss collocation methods (Trixi.jl, \codeinline{GaussSBP}, which uses affine mappings,
and FLUXO, \codeinline{GaussSBP}, which uses curvilinear mappings)
in both two and three dimensions. We observe that the Legendre-Gauss collocation schemes are
between \SI{1.5}{\times} and \SI{5}{\times}
more expensive than LGL-DGSEM, with the gap in performance closing as the order of
approximation increases in three dimensions. We note that these results imply
that the estimate of the cost of Legendre-Gauss collocation in \cite{chan2019efficient} was
optimistic\footnote{In \cite{chan2019efficient}, it was shown that the number of flux evaluations
for a degree $\polydeg$ entropy-conservative Legendre-Gauss scheme is slightly less than the number
of flux evaluations required for a degree $(\polydeg + 1)$ entropy-conservative LGL-DGSEM method.
Since the cost for entropy-conservative flux differencing schemes is typically dominated
by flux evaluations, it was argued that a degree $\polydeg$ entropy-conservative Legendre-Gauss scheme
will be roughly the cost of a degree $(\polydeg + 1)$ entropy-conservative LGL-DGSEM method.}, as
it did not take into account additional steps such as the entropy projection, interpolation of solution values to face nodes,
and lifting of face contributions to volume nodes.

Finally, we note that the
implementation of entropy-conservative Legendre-Gauss collocation schemes in Trixi.jl are slightly
less optimized compared to the implementation of LGL-DGSEM schemes.
First, for compatibility with analysis and visualization routines, Gauss schemes in
Trixi.jl store the solution at Legendre-Gauss-Lobatto nodes and interpolate to Legendre-Gauss nodes prior to
each right hand side evaluation. This interpolation can be performed in an efficient
tensor product fashion, though this does still result in some overhead. The second
difference is that Legendre-Gauss schemes compute surface fluxes locally on each element,
as done in \cite{hesthaven2007nodal}, which results in fluxes being computed twice
per face. In contrast, the implementation of LGL-DGSEM in Trixi.jl computes surface fluxes
only once per face then passes the computed fluxes to adjacent elements.
Based on the \PID results reported in Figure~\ref{fig:GaussPID}, however, we do not
expect these implementational differences to drastically change the overall runtime of
Legendre-Gauss collocation methods.

\section{More invasive optimizations}
\label{sec:invasive}

The optimizations discussed hitherto are noninvasive in the sense that they can
be applied to any flux differencing implementations. 
Sometimes, it can be beneficial to specialize the implementations even
further for specific conservation laws --- often at the cost of increased code
complexity. We discuss precomputing primitive variables and certain
logarithms needed in EC fluxes for the compressible Euler equations. Moreover,
we present some technical optimization techniques such as inlining volume
fluxes explicitly. All numerical experiments in this section are based on
LGL-DGSEM implementations in \trixi and \fluxo.

\subsection{Precomputing primitive variables for the compressible Euler equations}
\label{sec:primitive_variables}

Here, we benchmark the performance gains of volume terms by precomputing
the primitive variables $(\rho, v, p)$ for the compressible Euler equations
\eqref{eq:Euler}. All implementation techniques discussed so far can of course
be used to improve the efficiency of flux differencing algorithms using the
conservative variables $u = (\rho, \rho v, \rho e)$ directly. However, several
divisions and other arithmetic operations necessary to compute the primitive
variables from the conservative variables can be saved by precomputing them.
For this task, we use efficient specialized implementations based on
LoopVectorization.jl.

The benchmark setup is the same as in Section~\ref{sec:overintegration}, \ie,
we benchmark the total runtime of the volume term computation for a single
element on the Cartesian \codeinline{TreeMesh} of \trixi initialized with the
isentropic vortex initial condition \eqref{eq:isentropic-vortex}.

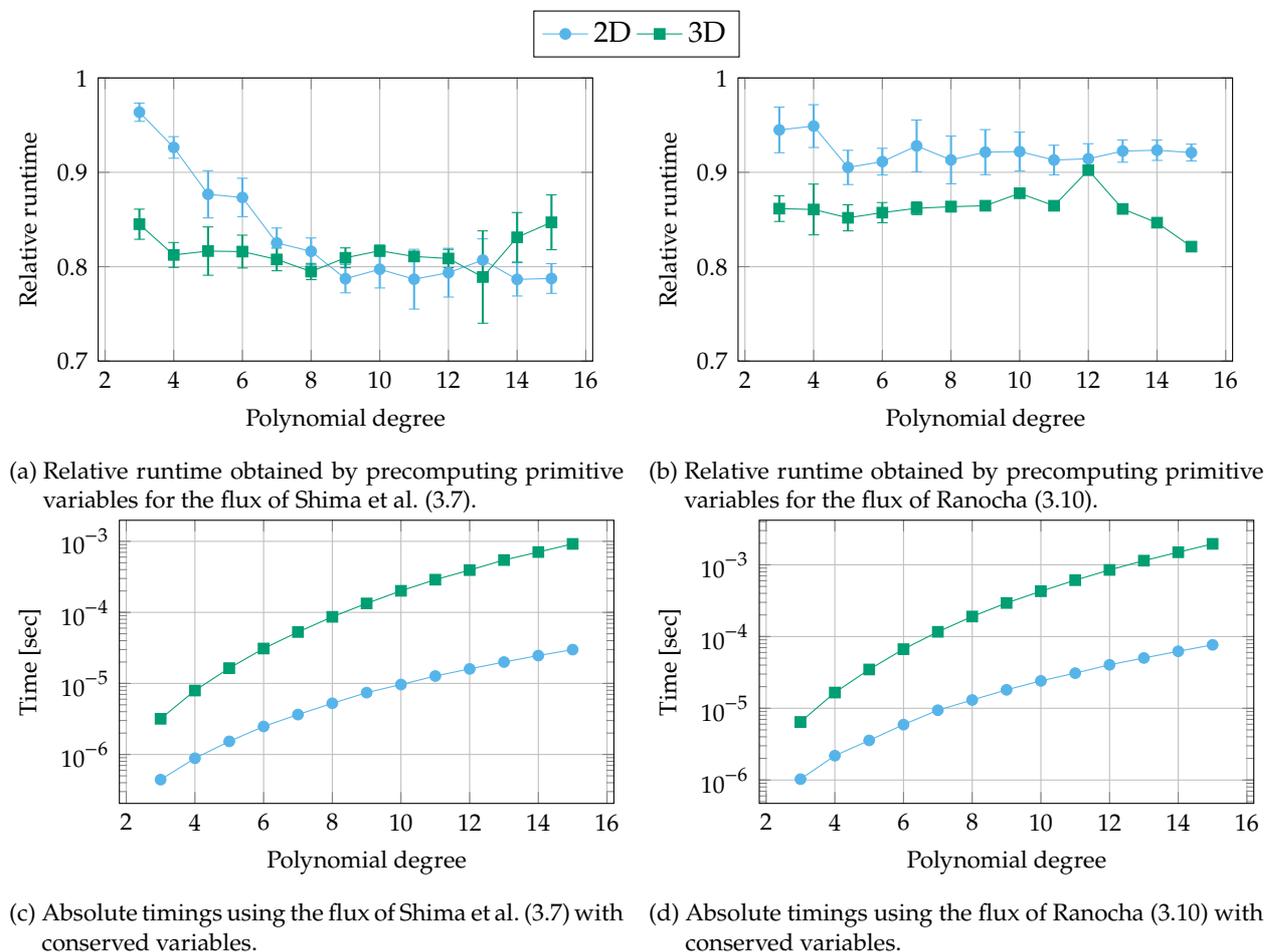
\begin{figure}[!htb]
\centering
  \pgfplotslegendfromname{legend:primitive_variables}
  \\
  \begin{subfigure}{0.49\linewidth}
    \begin{tikzpicture}[
      font=\footnotesize
      ]
      \begin{axis}[
          xlabel={Polynomial degree},
          ylabel={Relative runtime},
          width=\textwidth,
          height=0.66\textwidth,
          legend columns=-1,
          legend to name=legend:primitive_variables,
          ymin=0.7,
          ymax=1.00,
          grid=major,
        ]
        \addplot+ [error bars/.cd, y dir=both,y explicit,]
          table [x index=0, y index=1, y error index=2]{code/primitive_variables/primitive_2D_flux_shima_etal_relative.dat};
          \addlegendentry{2D}

        \addplot+ [error bars/.cd, y dir=both,y explicit,]
          table [x index=0, y index=1, y error index=2]{code/primitive_variables/primitive_3D_flux_shima_etal_relative.dat};
          \addlegendentry{3D}
      \end{axis}
    \end{tikzpicture}%
    \caption{Relative runtime obtained by precomputing primitive variables
             for the flux of Shima \etal \eqref{eq:flux_shima_etal}.}
  \end{subfigure}%
  \hspace*{\fill}
  \begin{subfigure}{0.49\linewidth}
    \begin{tikzpicture}[
      font=\footnotesize
      ]
      \begin{axis}[
          xlabel={Polynomial degree},
          ylabel={Relative runtime},
          width=\textwidth,
          height=0.66\textwidth,
          ymin=0.7,
          ymax=1.00,
          grid=major,
        ]
        \addplot+ [error bars/.cd, y dir=both,y explicit,]
          table [x index=0, y index=1, y error index=2]{code/primitive_variables/primitive_2D_flux_ranocha_relative.dat};

        \addplot+ [error bars/.cd, y dir=both,y explicit,]
          table [x index=0, y index=1, y error index=2]{code/primitive_variables/primitive_3D_flux_ranocha_relative.dat};
      \end{axis}
    \end{tikzpicture}%
    \caption{Relative runtime obtained by precomputing primitive variables
             for the flux of Ranocha \eqref{eq:flux_ranocha}.}
  \end{subfigure}%
  \\
  \begin{subfigure}{0.49\linewidth}
    \begin{tikzpicture}[
      font=\footnotesize
      ]
      \begin{semilogyaxis}[
          xlabel={Polynomial degree},
          ylabel={Time {[sec]}},
          width=\textwidth,
          height=0.66\textwidth,
          grid=major,
        ]
        \addplot+ [error bars/.cd, y dir=both,y explicit,]
          table [x index=0, y index=1, y error index=2]{code/primitive_variables/primitive_2D_flux_shima_etal.dat};

        \addplot+ [error bars/.cd, y dir=both,y explicit,]
          table [x index=0, y index=1, y error index=2]{code/primitive_variables/primitive_3D_flux_shima_etal.dat};
      \end{semilogyaxis}
    \end{tikzpicture}%
    \caption{Absolute timings using the flux of Shima \etal \eqref{eq:flux_shima_etal}
             with conserved variables.}
  \end{subfigure}%
  \hspace*{\fill}
  \begin{subfigure}{0.49\linewidth}
    \begin{tikzpicture}[
      font=\footnotesize
      ]
      \begin{semilogyaxis}[
          xlabel={Polynomial degree},
          ylabel={Time {[sec]}},
          width=\textwidth,
          height=0.66\textwidth,
          grid=major,
        ]
        \addplot+ [error bars/.cd, y dir=both,y explicit,]
          table [x index=0, y index=1, y error index=2]{code/primitive_variables/primitive_2D_flux_ranocha.dat};

        \addplot+ [error bars/.cd, y dir=both,y explicit,]
          table [x index=0, y index=1, y error index=2]{code/primitive_variables/primitive_3D_flux_ranocha.dat};
      \end{semilogyaxis}
    \end{tikzpicture}%
    \caption{Absolute timings using the flux of Ranocha \eqref{eq:flux_ranocha}
             with conserved variables.}
  \end{subfigure}%
  \caption{Microbenchmarks of precomputing the primitive variables compared to using
           the conserved variables (baseline) for flux differencing volume terms
           of LGL-DGSEM discretizations with polynomials of degree $\polydeg$ for
           the compressible Euler equations.}
  \label{fig:primitive_variables}
\end{figure}

The results of this comparison are shown in Figure~\ref{fig:primitive_variables}.
Clearly, precomputing the primitive variables improves the efficiency of the
volume term computations significantly, approximately between \SI{5}{\percent}
and \SI{25}{\percent} (on this computer system).
Usually, 3D computations benefit more from this invasive optimization than
2D computations. Moreover, relatively cheap numerical fluxes such as the one of
Shima \etal \eqref{eq:flux_shima_etal} benefit more than relatively expensive
EC fluxes.

These runtime improvements come at the cost of additional memory
requirements. Naively computing primitive variables at all volume nodes on a
Cartesian mesh in 3D requires additional memory $5 \cdot \ndofs$. Since any time
integration method will also require at least the same amount of temporary storage,
this would at most increase the memory requirements by one half. However, a
curved mesh requires storing the curvilinear coordinate information (contravariant
basis vectors), which requires $9 \cdot \ndofs$ memory. Thus, the additional
storage requirement is at most approx.\ one quarter. This can be further reduced by
computing and storing the primitive variables only for a single element before
computing the corresponding volume terms.

\subsection{Precomputing logarithms for the compressible Euler equations}
\label{sec:precomputing-log}

In addition to precomputing the primitive variables, one may also be interested in
computing other auxiliary quantities used in the evaluation of numerical fluxes. For
example, the evaluation of EC fluxes for the compressible Euler
equations requires computing the logarithmic mean of density and pressure, which
in turn requires computing the natural logarithm of density and pressure at two sets
of solution states. Computing logarithms (and other special functions) is significantly
more expensive than performing basic arithmetic operations. For example, on a 2019
Macbook Pro laptop (with a 2.3 GHz Intel\textregistered\ Core\texttrademark\ i9 processor), computing
$\log(x) + \log(y)$ takes between \SI{11.5 \pm 3.8}{\ns}, while computing a
simpler arithmetic operation such as $2x + 3y$ takes \SI{1.7 \pm 0.4}{\ns}.

For a degree $\polydeg$ approximation on a single tensor product element in $\ndims$
dimensions, computing entropy-conservative two-point fluxes for the compressible
Euler equations requires $\O\bigl(\ndims (\polydeg + 1)^{\ndims + 1} \bigr)$
logarithm evaluations. Precomputing logarithms for
density and pressure reduces this to $\O\bigl( (\polydeg + 1)^{\ndims} \bigr)$ evaluations, as
logarithms are computed once for each solution node instead of once for each pair of
nodes. We perform microbenchmarks for the entropy-conservative flux \eqref{eq:flux_ranocha}
similar to those in Section~\ref{sec:primitive_variables}. Specifically, we precompute
logarithms of density and pressure in addition to precomputing the primitive variables.

\begin{figure}[!htb]
\centering
  \pgfplotslegendfromname{legend:precomputed_logs}
  \\
  \begin{tikzpicture}[
    font=\footnotesize
    ]
    \begin{axis}[
        xlabel={Polynomial degree},
        ylabel={Relative change in runtime},
        width=0.75\textwidth,
        height=0.5\textwidth,
        legend columns=-1,
        legend to name=legend:precomputed_logs,
        ymin=0.2,
        ymax=1.2,
        grid=major,
      ]
      \node at (axis cs:10,1) [anchor=south] {isentropic vortex initial condition};
      \addplot+ [error bars/.cd, y dir=both,y explicit]
        table [x index=0, y index=1, y error index=2]{code/precompute_logs/precomputed_2D_flux_ranocha_vortex_relative.dat};
        \addlegendentry{2D}

      \addplot+ [error bars/.cd, y dir=both,y explicit]
        table [x index=0, y index=1, y error index=2]{code/precompute_logs/precomputed_3D_flux_ranocha_vortex_relative.dat};
        \addlegendentry{3D}

      \pgfplotsset{cycle list shift=-2}
      \node at (axis cs:12.6,0.48) [anchor=south] {sinusoidal initial condition};
      \addplot+ [error bars/.cd, y dir=both,y explicit]
        table [x index=0, y index=1, y error index=2]{code/precompute_logs/precomputed_2D_flux_ranocha_sine_relative.dat};

      \addplot+ [error bars/.cd, y dir=both,y explicit]
        table [x index=0, y index=1, y error index=2]{code/precompute_logs/precomputed_3D_flux_ranocha_sine_relative.dat};

      \pgfplotsset{cycle list shift=-4}
      \node at (axis cs:10,0.22) [anchor=south] {random initial condition};
      \addplot+ [error bars/.cd, y dir=both,y explicit]
        table [x index=0, y index=1, y error index=2]{code/precompute_logs/precomputed_2D_flux_ranocha_random_relative.dat};

      \addplot+ [error bars/.cd, y dir=both,y explicit]
        table [x index=0, y index=1, y error index=2]{code/precompute_logs/precomputed_3D_flux_ranocha_random_relative.dat};
    \end{axis}
  \end{tikzpicture}%
  \caption{Relative runtime when precomputing both logarithms and primitive variables
           compared to precomputing only primitive variables (baseline) for flux
           differencing volume terms of LGL-DGSEM discretizations using the flux of
           Ranocha \eqref{eq:flux_ranocha} with polynomials of degree $\polydeg$
           for the compressible Euler equations.}
  \label{fig:precomputed_logs}
\end{figure}
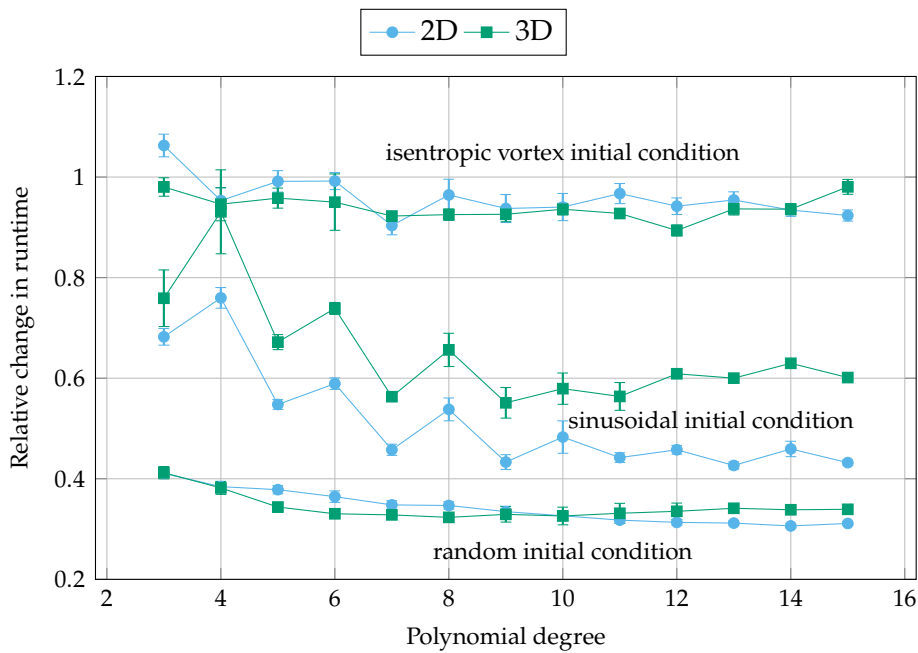

Figure~\ref{fig:precomputed_logs} shows relative runtimes when precomputing both
primitive variables and logarithms (compared with precomputing only the primitive
variables). We tested three initial conditions: the isentropic vortex \eqref{eq:isentropic-vortex}, a random
initial condition, and a sinusoidal initial condition
\begin{equation}
\begin{aligned}
\rho(x,y,0) &= 2 + \sin(\pi x / 5) \sin(\pi y / 5), && \text{(in 2D)}\\
\rho(x,y,z,0) &= 2 + \sin(\pi x / 5) \sin(\pi y / 5)\sin(\pi z / 5), && \text{(in 3D)}\\
p &= \rho^\gamma.
\end{aligned}
\end{equation}

For the isentropic vortex, we observe relatively modest speedup of roughly
\SI{10}{\percent}. In contrast, we observe more significant speedups
for the sinusoidal initial condition
(between \SI{25}{\percent} and \SI{55}{\percent})
and for the random initial condition
(between \SI{60}{\percent} and \SI{75}{\percent}).
This is because Algorithms~\ref{alg:logmean-optimized} and
\ref{alg:inv_logmean-optimized} for computing logarithmic means do not evaluate logarithms
between two solution states if the density or pressure are close to each other (instead, a
high order Taylor approximation is evaluated). Thus, precomputing logarithms does not yield
any speedup (and in fact is slightly slower) when the solution is near-constant.

For the isentropic vortex initial condition, the solution is compactly supported and thus
constant in a large percentage of the domain (especially in 3D). As a result, we observe
very modest gains in the overall runtime. For the sinusoidal and random initial conditions,
the solution varies globally over the entire domain, and we observe more significant speedup
when precomputing logarithms.

\subsection{Defining options at compile time}
\label{sec:compile-time}

The techniques discussed so far are at the level of the mathematical description
of the algorithms. While we will mostly stay at this high level in this article,
we would like to point out that an efficient implementation will also require
appropriate programming techniques. The effort required to achieve
these goals depends on the specific programming language. Here, we briefly
investigate the use of inlining of the volume fluxes and setting the polynomial
degree at compile time in \fluxo.

In traditional scientific computing languages such as Fortran, C, and C++,
the code is compiled prior to execution.
To avoid users having to modify and/or compile the source code multiple times
to change simulation options (such as the polynomial degree or the volume flux),
a common practice is to read all simulation parameters from a text file and assign
them at runtime.

The definition of functions used many times throughout the simulation at runtime
implies the repeated evaluation of switch statements during the computation or
the use of procedure pointers or type polymorphism.
The repeated evaluation of switch statements might increase code complexity and
deteriorate performance.
Moreover, even though procedure pointers and polymorphic types can be used
to simplify code, many compilers and their branch predictors fail at inlining
functions that use them, which also affects performance.

\fluxo allows the user to specify the polynomial degree and the volume flux
at runtime for flexibility, but also provides the possibility to define these
and other options at compile time to improve performance of long runs.
Of course, the coexistence of compile time and runtime specifications increases
code complexity.

\begin{figure}[!htb]
\centering
  \pgfplotslegendfromname{legend:Compile}
  \\
  \begin{tikzpicture}[
    font=\footnotesize
    ]
    \begin{axis}[
        xlabel={Polynomial degree},
        ylabel={Time/RHS/DOF {[sec]}},
        width=0.66\textwidth,
        height=0.44\textwidth,
        legend columns=3,
        legend to name=legend:Compile,
        grid=major,
      ]
      \addplot+ [error bars/.cd, y dir=both,y explicit]
        table [x index=0, y index=1, y error index=2, col sep=comma]{code/ec_performance_fluxo/serialTests/results/results_Ranocha_noN_noFlux.dat};
        \addlegendentry{$\polydeg$ (r); volume flux (r)}

      \addplot+ [error bars/.cd, y dir=both,y explicit]
        table [x index=0, y index=1, y error index=2, col sep=comma]{code/ec_performance_fluxo/serialTests/results/results_Ranocha_noN.dat};
        \addlegendentry{$\polydeg$ (r); volume flux (c)}

      \addplot+ [error bars/.cd, y dir=both,y explicit]
        table [x index=0, y index=1, y error index=2, col sep=comma]{code/ec_performance_fluxo/serialTests/results/results_Ranocha.dat};
        \addlegendentry{$\polydeg$ (c); volume flux (c)}
    \end{axis}
  \end{tikzpicture}%
  \caption{Influence of setting the volume flux and the polynomial degree $\polydeg$
           at compile time~(c) or at runtime~(r) on the \PID of \fluxo for
           entropy-conservative LGL-DGSEM discretizations using the flux of Ranocha
           \eqref{eq:flux_ranocha} for the compressible Euler equations.}
  \label{fig:Compile}
\end{figure}
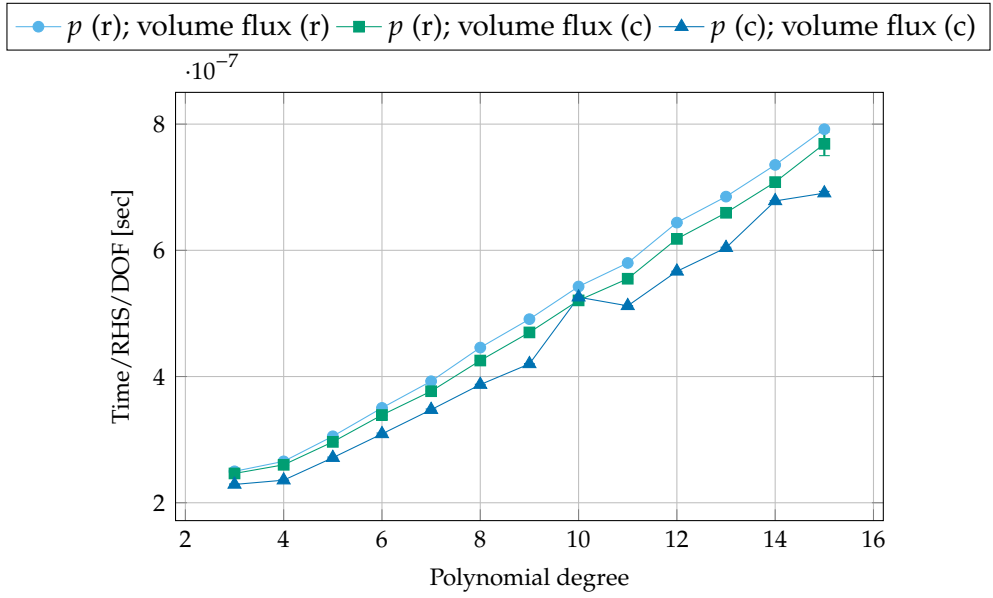

Figure~\ref{fig:Compile} shows computed \PID values obtained with
LGL-DGSEM of \fluxo for the isentropic vortex setup introduced in
Section~\ref{sec:Cartesian-vs-curved}.
Setting the volume flux at compile time enables more compiler optimizations
such as inlining and thus increases the performance by approx.\ \SI{5}{\percent}.
Setting also the polynomial degree at compile time options allows more
optimizations, resulting in a speedup of up to \SI{15}{\percent}.

Julia \cite{bezanson2017julia} uses a ``just ahead of time'' (or ``just in time'')
compiler approach. In particular, user defined functions can be inlined into
library code without additional programming effort. Currently, the
Julia compiler uses a heuristic to determine whether a function should be inlined.
Programmers can hint the compiler to inline a function by prepending the function
definition by the macro \codeinline{@inline}.

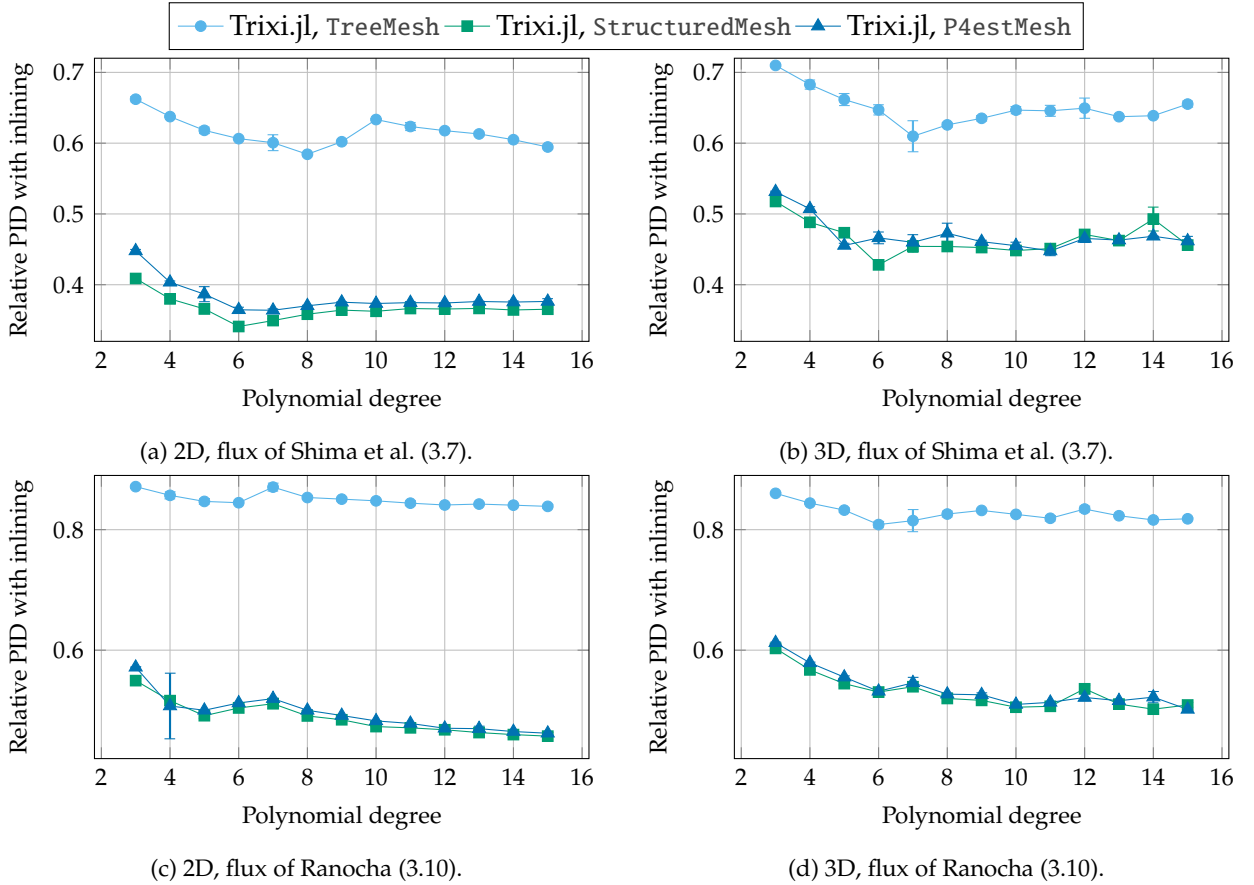
\begin{figure}[!htb]
\centering
  \pgfplotslegendfromname{legend:inlining}
  \\
  \begin{subfigure}{0.49\linewidth}
    \begin{tikzpicture}[
      font=\footnotesize
      ]
      \begin{axis}[
          xlabel={Polynomial degree},
          ylabel={Relative \PID with inlining},
          width=\textwidth,
          height=0.66\textwidth,
          ymin=0.32,
          ymax=0.72,
          grid=major,
        ]
        \addplot+ [error bars/.cd, y dir=both,y explicit,]
          table [x index=0, y index=1, y error index=2]{code/inlining/pids_2D_flux_shima_etal_notinlined_relative.dat};

        \addplot+ [error bars/.cd, y dir=both,y explicit,]
          table [x index=0, y index=3, y error index=4]{code/inlining/pids_2D_flux_shima_etal_notinlined_relative.dat};

        \addplot+ [error bars/.cd, y dir=both,y explicit,]
          table [x index=0, y index=5, y error index=6]{code/inlining/pids_2D_flux_shima_etal_notinlined_relative.dat};
      \end{axis}
    \end{tikzpicture}%
    \caption{2D, flux of Shima \etal \eqref{eq:flux_shima_etal}.}
  \end{subfigure}%
  \hspace*{\fill}
  \begin{subfigure}{0.49\linewidth}
    \begin{tikzpicture}[
      font=\footnotesize
      ]
      \begin{axis}[
          xlabel={Polynomial degree},
          ylabel={Relative \PID with inlining},
          width=\textwidth,
          height=0.66\textwidth,
          ymin=0.32,
          ymax=0.72,
          grid=major,
        ]
        \addplot+ [error bars/.cd, y dir=both,y explicit,]
          table [x index=0, y index=1, y error index=2]{code/inlining/pids_3D_flux_shima_etal_notinlined_relative.dat};

        \addplot+ [error bars/.cd, y dir=both,y explicit,]
          table [x index=0, y index=3, y error index=4]{code/inlining/pids_3D_flux_shima_etal_notinlined_relative.dat};

        \addplot+ [error bars/.cd, y dir=both,y explicit,]
          table [x index=0, y index=5, y error index=6]{code/inlining/pids_3D_flux_shima_etal_notinlined_relative.dat};
      \end{axis}
    \end{tikzpicture}%
    \caption{3D, flux of Shima \etal \eqref{eq:flux_shima_etal}.}
  \end{subfigure}%
  \\
  \begin{subfigure}{0.49\linewidth}
    \begin{tikzpicture}[
      font=\footnotesize
      ]
      \begin{axis}[
          xlabel={Polynomial degree},
          ylabel={Relative \PID with inlining},
          width=\textwidth,
          height=0.66\textwidth,
          ymin=0.42,
          ymax=0.89,
          grid=major,
        ]
        \addplot+ [error bars/.cd, y dir=both,y explicit,]
          table [x index=0, y index=1, y error index=2]{code/inlining/pids_2D_flux_ranocha_notinlined_relative.dat};

        \addplot+ [error bars/.cd, y dir=both,y explicit,]
          table [x index=0, y index=3, y error index=4]{code/inlining/pids_2D_flux_ranocha_notinlined_relative.dat};

        \addplot+ [error bars/.cd, y dir=both,y explicit,]
          table [x index=0, y index=5, y error index=6]{code/inlining/pids_2D_flux_ranocha_notinlined_relative.dat};
      \end{axis}
    \end{tikzpicture}%
    \caption{2D, flux of Ranocha \eqref{eq:flux_ranocha}.}
  \end{subfigure}%
  \hspace*{\fill}
  \begin{subfigure}{0.49\linewidth}
    \begin{tikzpicture}[
      font=\footnotesize
      ]
      \begin{axis}[
          xlabel={Polynomial degree},
          ylabel={Relative \PID with inlining},
          width=\textwidth,
          height=0.66\textwidth,
          legend to name=legend:inlining,
          legend columns=-1,
          ymin=0.42,
          ymax=0.89,
          grid=major,
        ]
        \addplot+ [error bars/.cd, y dir=both,y explicit,]
          table [x index=0, y index=1, y error index=2]{code/inlining/pids_3D_flux_ranocha_notinlined_relative.dat};
          \addlegendentry{\trixi, \codeinline{TreeMesh}}

        \addplot+ [error bars/.cd, y dir=both,y explicit,]
          table [x index=0, y index=3, y error index=4]{code/inlining/pids_3D_flux_ranocha_notinlined_relative.dat};
          \addlegendentry{\trixi, \codeinline{StructuredMesh}}

        \addplot+ [error bars/.cd, y dir=both,y explicit,]
          table [x index=0, y index=5, y error index=6]{code/inlining/pids_3D_flux_ranocha_notinlined_relative.dat};
          \addlegendentry{\trixi, \codeinline{P4estMesh}}
      \end{axis}
    \end{tikzpicture}%
    \caption{3D, flux of Ranocha \eqref{eq:flux_ranocha}.}
  \end{subfigure}%
  \caption{Relative performance obtained by inlining the volume fluxes for flux
           differencing volume terms of LGL-DGSEM discretizations with polynomials of
           degree $\polydeg$ for the compressible Euler equations. The absolute
           \PID with inlining is visualized in Figure~\ref{fig:Cartesian-vs-curved-PID-Euler}.}
  \label{fig:inlining}
\end{figure}

Figure~\ref{fig:inlining} shows the effect of inlining volume fluxes in Trixi.jl.
As expected, the relative improvement of this optimization is better for cheap
volume fluxes such as the one of Shima \etal \eqref{eq:flux_shima_etal}.
Additionally, inlining is more important on the curved meshes of Trixi.jl.
There, it can reduce the \PID by up to approx.\ \SI{2}{\times} (compared to
approx.\ \SI{25}{\percent} on the Cartesian \codeinline{TreeMesh}).

\section{Explicit SIMD optimizations}
\label{sec:SIMD}

The operational intensity of flux differencing volume terms is relatively high.
Thus, we expect to be in the compute bound regime in a standard roofline
model \cite{williams2009roofline} of standard CPUs. This is the reason why
we described several optimizations that reduce the amount of work and increase
runtime performance. To further increase the (serial) performance, instruction
level parallelization is required.
To make optimal use of modern CPUs, utilizing SIMD (single instruction, multiple data) instructions is key. However,
current compiler and code generation techniques are often not advanced enough
to handle all expressions efficiently. Thus, some manual intervention is needed
to write SIMD-friendly code. Here, we focus on implementations based on
LoopVectorization.jl in Trixi.jl. The same kind of optimizations are also effective
for auto-vectorization by decent Fortran compilers in FLUXO.

Trixi.jl is written as a research code accessible to students and newcomers.
In particular, some design choices are based on the goal to make setting up
new physical models easy. Thus, physics including (numerical) fluxes is handled
pointwise by small (inlined) functions and the global solution uses an array
of structures (AoS) memory layout. To enable efficient SIMD optimizations at the
element level, we permute array dimensions effectively to an array of structures
of arrays (AoSoA), i.e., we use a temporary structure of arrays for a single element.
Loops in a tensor product ansatz are structured to keep the triangular part
at the outermost loop (making use of the skew-symmetry of the flux differencing
operator \eqref{eq:opDsplit})
and plain loops over all nodes in the inner part for SIMD
vectorization. Moreover, the memory is rearranged to ensure the first
dimension is one of the dimensions to which SIMD instructions can be applied (since Julia
uses a Fortran-style column-major memory layout by default). Such a rearrangement of memory
is also crucial for Fortran compilers to optimize similar parts
in FLUXO \cite{ribeiro2020optdg2}.

Moreover, we precompute primitive variables before computing the flux
differencing volume terms (\cf Section~\ref{sec:primitive_variables}).
For the EC flux requiring logarithmic means, we also precompute logarithms
of the density $\rho$ and the pressure $p$. In contrast to the results reported
in Section~\ref{sec:precomputing-log}, this step becomes more important even for
the isentropic vortex initial condition.
Scalar logarithm implementations are often based on algorithms including table
lookups and conditional branches, e.g., \cite{tang1990table}. In contrast,
logarithm implementations optimized for SIMD instructions are usually more
demanding and cannot use fast paths. We observed up to ca. \SI{2}{\times} speed-up by
precomputing logarithms on an Intel\textregistered\ Core\texttrademark\ i7-8700K (CPU from 2017 with AVX2)
for the isentropic vortex initial condition.

All of these specializations are available for the fluxes of
Shima \etal \eqref{eq:flux_shima_etal} and Ranocha \eqref{eq:flux_ranocha}
on 2D and 3D meshes in recent versions of Trixi.jl. The benchmarks reported thus far
do not use these SIMD optimizations.

\begin{table}[htb]
\centering
\caption{Performance metrics of 3D flux differencing volume terms in Trixi.jl
         using SIMD optimizations on an Intel\textregistered\ Core\texttrademark\ i7-8700K (CPU from 2017 with AVX2).}
\label{tab:peak_performance}
\begin{tabular*}{\linewidth}{@{\extracolsep{\fill}} l SSS @{}}
  \toprule
    & \multicolumn{1}{c}{\texttt{TreeMesh}}
    & \multicolumn{1}{c}{\texttt{StructuredMesh}}
    & \multicolumn{1}{c}{\texttt{P4estMesh}}
  \\
  \midrule
  Flux of Shima \etal \eqref{eq:flux_shima_etal}
  \\
  Vectorization ratio in \%
    & \num{98.36}
    & \num{98.85}
    & \num{98.85}
  \\
  Absolute performance in Gflops/s
    & \num{21.44}
    & \num{18.51}
    & \num{18.44}
  \\
  Relative to peak performance in \%
    & \num{29.09}
    & \num{25.12}
    & \num{25.02}
  \\
  \midrule
  Flux of Ranocha \eqref{eq:flux_ranocha}
  \\
  Vectorization ratio in \%
    & \num{99.09}
    & \num{99.26}
    & \num{99.26}
  \\
  Absolute performance in Gflops/s
    & \num{18.80}
    & \num{15.28}
    & \num{15.25}
  \\
  Relative to peak performance in \%
    & \num{25.51}
    & \num{20.74}
    & \num{20.69}
  \\
  \bottomrule
\end{tabular*}
\end{table}

We used LIKWID \cite{treibig2010likwid} via its Julia interface LIKWID.jl
to measure some performance metrics of
the new volume terms optimized for SIMD instructions. For this, we used the same
setup as in Section~\ref{sec:overintegration} with polynomials of degree
$\polydeg = 3$. We used a single batch computing the volume terms $5 \cdot 10^3$
times on an Intel\textregistered\ Core\texttrademark\ i7-8700K (CPU from 2017 with AVX2). The results are
shown in Table~\ref{tab:peak_performance}. For all numerical fluxes and meshes,
we achieved vectorization ratios of more than \SI{98}{\percent}. This includes all
required memory rearrangements described above.

We used the same setup with LIKWID to estimate the percentage of peak
performance of floating point operations per second.
We used the LIKWID benchmark tool to estimate the peak performance for fused
multiply-add (FMA) instructions and measured the floating point performance of
the flux differencing volume terms in Trixi.jl. The results are also shown in
Table~\ref{tab:peak_performance}.
Given that the estimated peak performance uses only FMAs and is rather optimistic
while the actual algorithm involves other operations including divisions
and logarithms for the EC flux, the obtained performance results are quite good.
The volume terms on the curved meshes require additional memory rearrangements
of the contravariant basis vectors (due to the choice of AoS style memory
layouts described above). Together with the increased memory loads, this explains
the reduced peak performance of flux differencing on curved meshes.

\begin{figure}[!htb]
\centering
  \begin{subfigure}{\linewidth}
  \centering
    \begin{tikzpicture}
        \begin{axis}[
            xlabel={Operational intensity in Flops/Byte},
            ylabel={Attainable Gflops/s},
            height=0.5\textwidth,
            width=\textwidth,
            xmin=0.9,
            xmax=3.9,
            ymin=3,  
            ymax=81, 
          ]
          %
          \addplot[mark=none, thick, black, samples=2, domain=0.125:3.356839224260235] {x * 21.95787109375};
          \addplot[mark=none, thick, black, samples=2, domain=3.356839224260235:5] {73.70904296875};

          \node[label={[label distance=-3pt]0:{\codeinline{TreeMesh}}}, circle, fill, inner sep=2pt] at (axis cs:2.3471140158878363, 8.512998787030106) {};

          \node[label={[label distance=-3pt]90:{\codeinline{StructuredMesh}}}, rectangle, fill, inner sep=2pt] at (axis cs:1.9135305573588055, 8.622553567902145) {};

          \node[label={[label distance=-3pt]2700:{\codeinline{P4estMesh}}}, shape=diamond, fill, inner sep=1.41pt] at (axis cs:1.8967146507822699, 8.486387884488854) {};

          \node[label={[label distance=-3pt]0:{\codeinline{TreeMesh}, SIMD}}, circle, fill, inner sep=2pt] at (axis cs:1.943287572537641, 21.289385660049444) {};

          \node[label={[label distance=-3pt, align=center]90:{\codeinline{StructuredMesh},\\SIMD}}, rectangle, fill, inner sep=2pt] at (axis cs:1.6191706769274423, 18.46107355065368) {};

          \node[label={[label distance=-4pt]225:{\codeinline{P4estMesh}, SIMD}}, shape=diamond, fill, inner sep=1.41pt] at (axis cs:1.6386543614291538, 18.401884419017208) {};
        \end{axis}
    \end{tikzpicture}
    \caption{Flux of Shima \etal \eqref{eq:flux_shima_etal}.}
  \end{subfigure}%
  \\
  \begin{subfigure}{\linewidth}
  \centering
    \begin{tikzpicture}
        \begin{axis}[
            xlabel={Operational intensity in Flops/Byte},
            ylabel={Attainable Gflops/s},
            height=0.5\textwidth,
            width=\textwidth,
            xmin=0.9,
            xmax=3.9,
            ymin=3,  
            ymax=81, 
          ]
          %
          \addplot[mark=none, thick, black, samples=2, domain=0.125:3.356839224260235] {x * 21.95787109375};
          \addplot[mark=none, thick, black, samples=2, domain=3.356839224260235:5] {73.70904296875};

          \node[label={[label distance=-3pt]90:{\codeinline{TreeMesh}}}, circle, fill, inner sep=2pt] at (axis cs:3.518739853744992, 6.480704688759926) {};

          \node[label={[label distance=-3pt]0:{\codeinline{StructuredMesh}}}, rectangle, fill, inner sep=2pt] at (axis cs:2.659019038251165, 6.083740645433487) {};

          \node[label={[label distance=-3pt]180:{\codeinline{P4estMesh}}}, shape=diamond, fill, inner sep=1.41pt] at (axis cs:2.65506503400861, 6.062575093179439) {};

          \node[label={[label distance=-3pt]90:{\codeinline{TreeMesh}, SIMD}}, circle, fill, inner sep=2pt] at (axis cs:3.4471617184253858, 18.452692983673145) {};

          \node[label={[label distance=-3pt]270:{\codeinline{StructuredMesh}, SIMD}}, rectangle, fill, inner sep=2pt] at (axis cs:2.503611048970959, 15.419931350917922) {};

          \node[label={[label distance=-3pt]90:{\codeinline{P4estMesh}, SIMD}}, shape=diamond, fill, inner sep=1.41pt] at (axis cs:2.623427940076148, 15.558971175082915) {};
        \end{axis}
    \end{tikzpicture}
    \caption{Flux of Ranocha \eqref{eq:flux_ranocha}.}
  \end{subfigure}%
  \caption{Empirical roofline model \cite{williams2009roofline} measured using
           LIKWID \cite{treibig2010likwid} for the flux differencing volume terms
           of the 3D compressible Euler equations using polynomials of degree
           $\polydeg = 3$ and 8 elements per coordinate direction on an
           Intel\textregistered\ Core\texttrademark\ i7-8700K (CPU from 2017 with AVX2).
           Results are shown on the \codeinline{TreeMesh} (circles),
           \codeinline{StructuredMesh} (squares), and the
           \codeinline{P4estMesh} (diamonds) of Trixi.jl.}
  \label{fig:SIMD-roofline}
\end{figure}
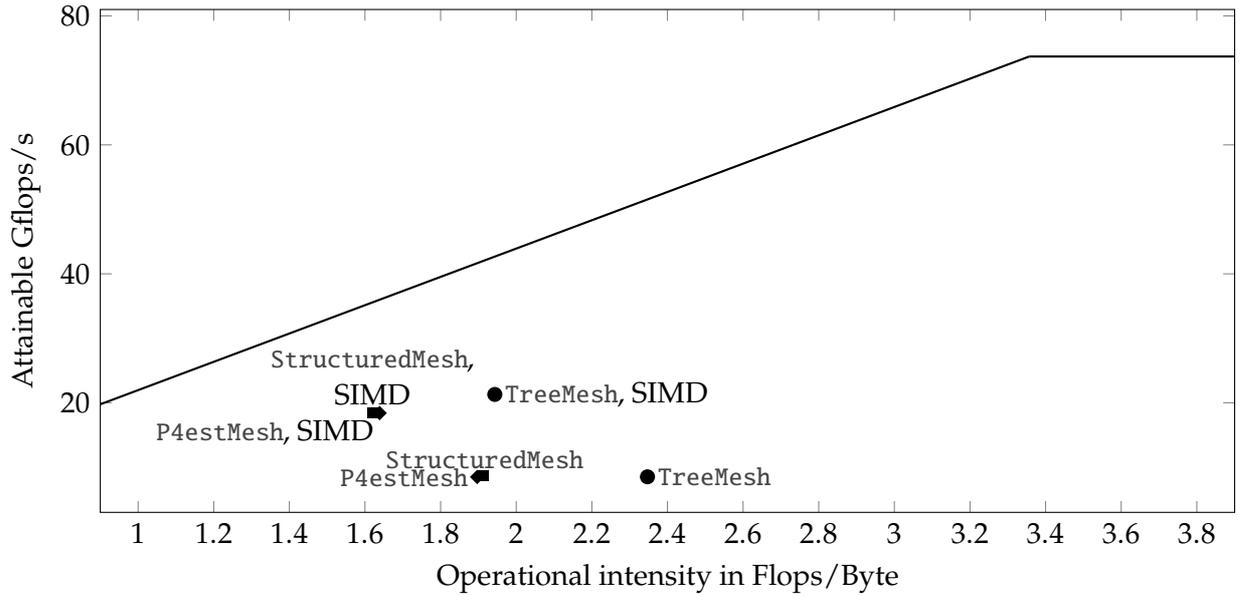
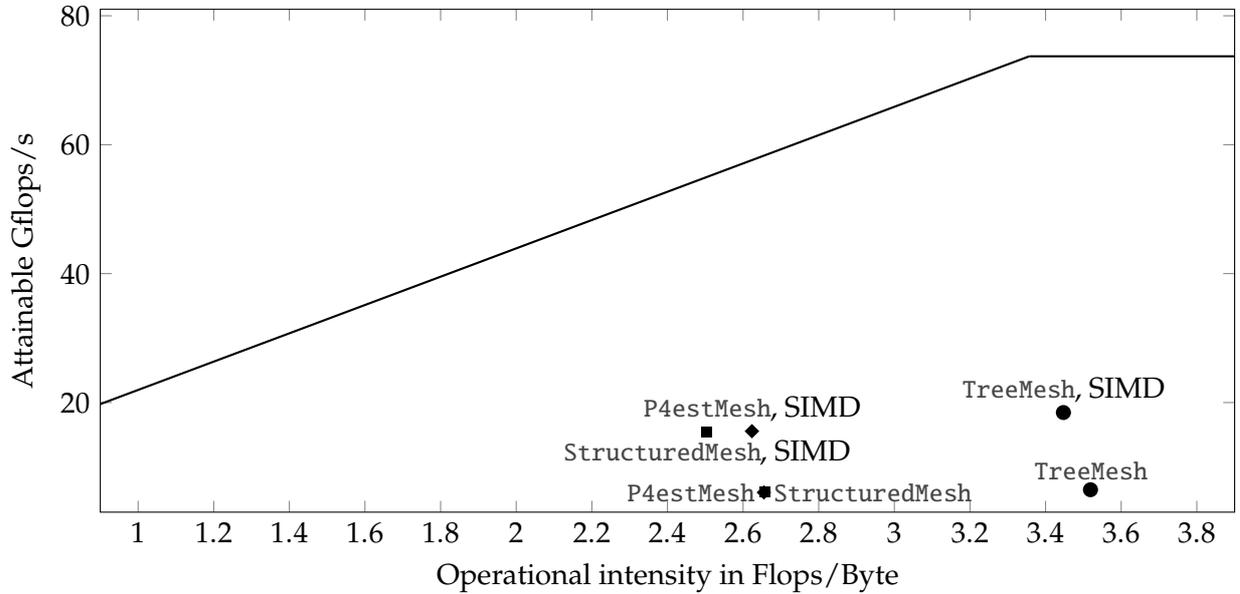

The results with and without SIMD optimizations described in this
section are embedded into an empirical roofline model shown in
Figure~\ref{fig:SIMD-roofline}. Again, the measurements have been
performed using LIKWID, following their tutorial on creating an
empirical roofline model. The peak performance has been estimated
using FMA instructions and the maximum bandwidth has been estimated
by load instructions (using LIKWID benchmark tools for both).
First, note that the volume terms with SIMD optimizations have a
reduced operational intensity compared to the plain versions. This is due
to the fact that we precompute primitive variables and logarithms.
Second, the SIMD-optimized versions have a higher performance.
On the system we used for this benchmark, many data points (in particular
for the cheaper non-EC flux) are under the slant of the roof, i.e.,
the regime that is usually characterized as limited by memory bandwidth.
However, note that the peak performance is estimated by pure FMA
instructions while the numerical fluxes involve more expensive
operations such as divisions.
Finally, these measurements are only valid when considering a single process
operating on the CPU. In case of a parallel simulation, all processes on a
single node need to share the same memory bandwidth.
This can affect the serial performance of each process and skew the
results further towards the memory-bound regime.

\begin{figure}[!htb]
\centering
  \pgfplotslegendfromname{legend:SIMD-PID-Euler}
  \\
  \begin{subfigure}{0.49\linewidth}
    \begin{tikzpicture}[
      font=\footnotesize
      ]
      \begin{axis}[
          xlabel={Polynomial degree},
          ylabel={Time/RHS/DOF {[sec]}},
          width=\textwidth,
          height=0.66\textwidth,
          ymin=0.0,
          grid=major,
        ]
        \addplot+ [error bars/.cd, y dir=both,y explicit,]
          table [x index=0, y index=1, y error index=2]{code/SIMD/pids_2D_flux_shima_etal_turbo.dat};

        \addplot+ [error bars/.cd, y dir=both,y explicit,]
          table [x index=0, y index=3, y error index=4]{code/SIMD/pids_2D_flux_shima_etal_turbo.dat};

        \addplot+ [error bars/.cd, y dir=both,y explicit,]
          table [x index=0, y index=5, y error index=6]{code/SIMD/pids_2D_flux_shima_etal_turbo.dat};
      \end{axis}
    \end{tikzpicture}%
    \caption{2D, flux of Shima \etal \eqref{eq:flux_shima_etal}.}
  \end{subfigure}%
  \hspace*{\fill}
  \begin{subfigure}{0.49\linewidth}
    \begin{tikzpicture}[
      font=\footnotesize
      ]
      \begin{axis}[
          xlabel={Polynomial degree},
          ylabel={Time/RHS/DOF {[sec]}},
          width=\textwidth,
          height=0.66\textwidth,
          ymin=0.0,
          grid=major,
        ]
        \addplot+ [error bars/.cd, y dir=both,y explicit,]
          table [x index=0, y index=1, y error index=2]{code/SIMD/pids_3D_flux_shima_etal_turbo.dat};

        \addplot+ [error bars/.cd, y dir=both,y explicit,]
          table [x index=0, y index=3, y error index=4]{code/SIMD/pids_3D_flux_shima_etal_turbo.dat};

        \addplot+ [error bars/.cd, y dir=both,y explicit,]
          table [x index=0, y index=5, y error index=6]{code/SIMD/pids_3D_flux_shima_etal_turbo.dat};
      \end{axis}
    \end{tikzpicture}%
    \caption{3D, flux of Shima \etal \eqref{eq:flux_shima_etal}.}
  \end{subfigure}%
  \\
  \begin{subfigure}{0.49\linewidth}
    \begin{tikzpicture}[
      font=\footnotesize
      ]
      \begin{axis}[
          xlabel={Polynomial degree},
          ylabel={Time/RHS/DOF {[sec]}},
          width=\textwidth,
          height=0.66\textwidth,
          ymin=0.0,
          grid=major,
        ]
        \addplot+ [error bars/.cd, y dir=both,y explicit,]
          table [x index=0, y index=1, y error index=2]{code/SIMD/pids_2D_flux_ranocha_turbo.dat};

        \addplot+ [error bars/.cd, y dir=both,y explicit,]
          table [x index=0, y index=3, y error index=4]{code/SIMD/pids_2D_flux_ranocha_turbo.dat};

        \addplot+ [error bars/.cd, y dir=both,y explicit,]
          table [x index=0, y index=5, y error index=6]{code/SIMD/pids_2D_flux_ranocha_turbo.dat};
      \end{axis}
    \end{tikzpicture}%
    \caption{2D, flux of Ranocha \eqref{eq:flux_ranocha}.}
  \end{subfigure}%
  \hspace*{\fill}
  \begin{subfigure}{0.49\linewidth}
    \begin{tikzpicture}[
      font=\footnotesize
      ]
      \begin{axis}[
          xlabel={Polynomial degree},
          ylabel={Time/RHS/DOF {[sec]}},
          width=\textwidth,
          height=0.66\textwidth,
          legend to name=legend:SIMD-PID-Euler,
          legend columns=-1,
          ymin=0.0,
          grid=major,
        ]
        \addplot+ [error bars/.cd, y dir=both,y explicit,]
          table [x index=0, y index=1, y error index=2]{code/SIMD/pids_3D_flux_ranocha_turbo.dat};
          \addlegendentry{\trixi, \codeinline{TreeMesh}}

        \addplot+ [error bars/.cd, y dir=both,y explicit,]
          table [x index=0, y index=3, y error index=4]{code/SIMD/pids_3D_flux_ranocha_turbo.dat};
          \addlegendentry{\trixi, \codeinline{StructuredMesh}}

        \addplot+ [error bars/.cd, y dir=both,y explicit,]
          table [x index=0, y index=5, y error index=6]{code/SIMD/pids_3D_flux_ranocha_turbo.dat};
          \addlegendentry{\trixi, \codeinline{P4estMesh}}
      \end{axis}
    \end{tikzpicture}%
    \caption{3D, flux of Ranocha \eqref{eq:flux_ranocha}.}
  \end{subfigure}%
  \caption{Runtime per right-hand side evaluation and degree of freedom for
           different mesh types and LGL-DGSEM discretizations
           with SIMD optimizations of the volume terms
           for the compressible Euler equations.}
  \label{fig:SIMD-PID-Euler}
\end{figure}
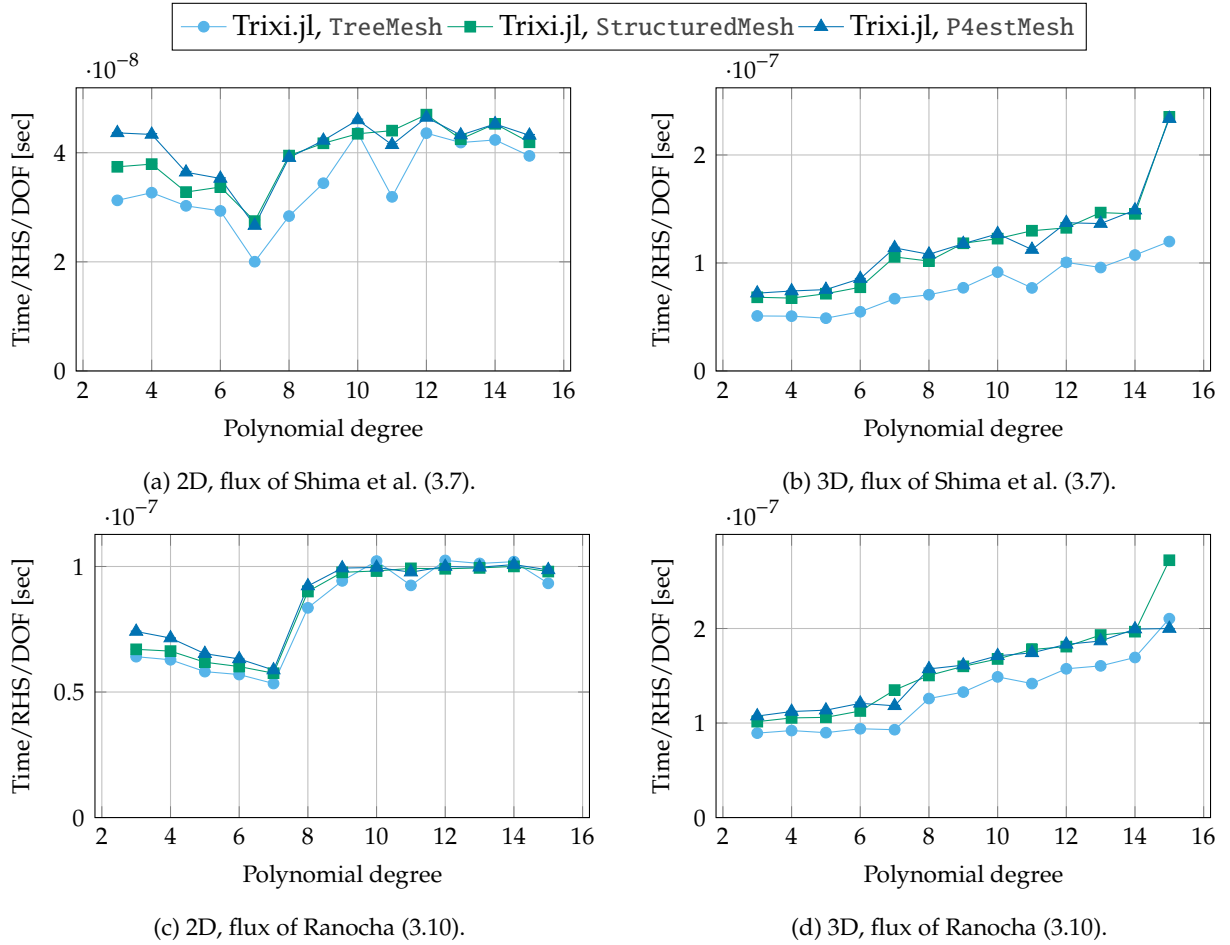

The new \PID measurements including precomputing terms and SIMD optimizations are shown in
Figure~\ref{fig:SIMD-PID-Euler}. In contrast to the baseline results shown in
Figure~\ref{fig:Cartesian-vs-curved-PID-Euler}, the {\PID} does not scale linearly
with the polynomial degree. In 2D, the \PID is approximately linearly
\emph{decreasing} for $\polydeg \in \{ 3, \dots, 7 \}$. After the local minimum
at $\polydeg = 7$, the \PID plateaus again at a higher level.
In 3D, a rough linear trend of the \PID can still be observed. However, the
\PID plateaus for small polynomial degrees up to $\polydeg = 6$ or $\polydeg = 7$,
depending on the mesh type and numerical flux.

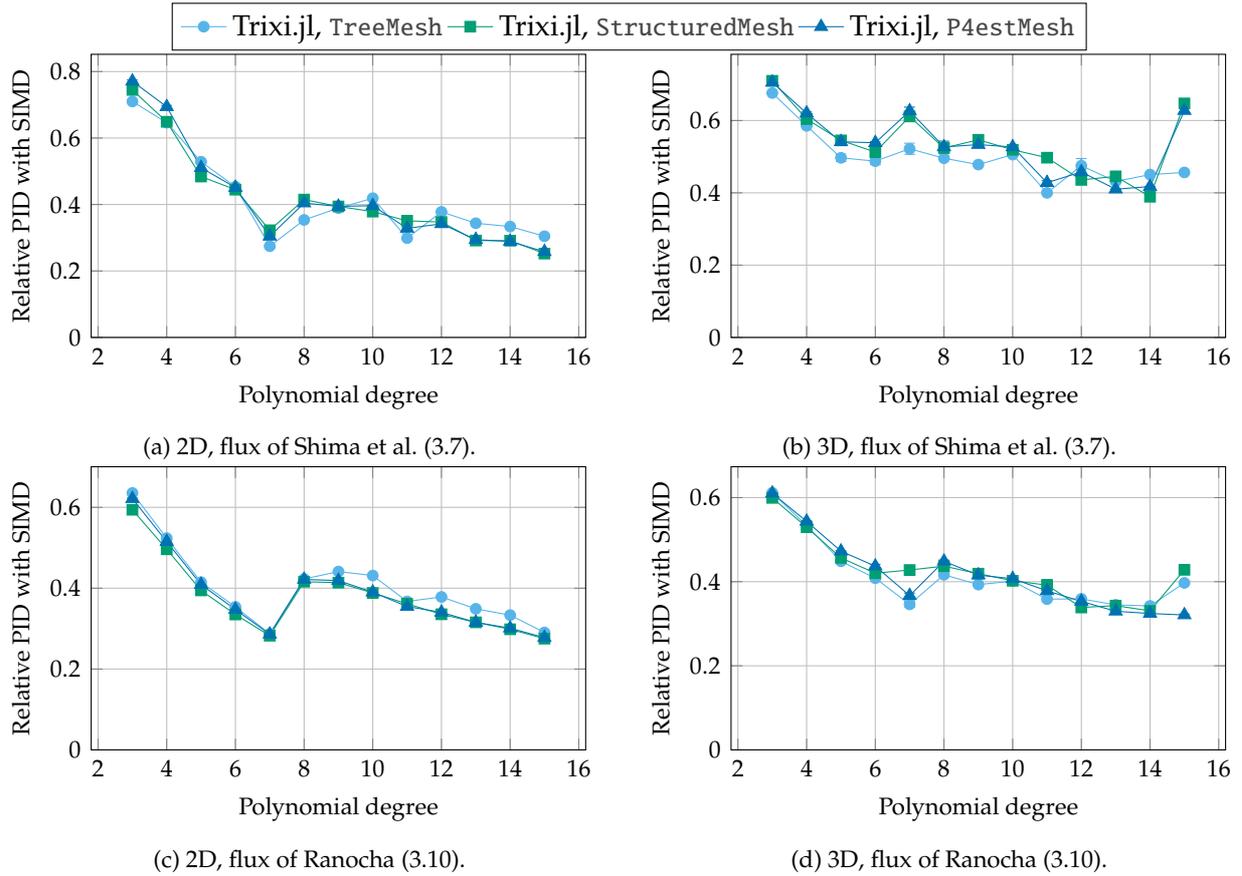
\begin{figure}[!htb]
\centering
  \pgfplotslegendfromname{legend:SIMD-relative-Euler}
  \\
  \begin{subfigure}{0.49\linewidth}
    \begin{tikzpicture}[
      font=\footnotesize
      ]
      \begin{axis}[
          xlabel={Polynomial degree},
          ylabel={Relative \PID with SIMD},
          width=\textwidth,
          height=0.66\textwidth,
          ymin=0.0,
          grid=major,
        ]
        \addplot+ [error bars/.cd, y dir=both,y explicit,]
          table [x index=0, y index=1, y error index=2]{code/SIMD/pids_2D_flux_shima_etal_turbo_relative.dat};

        \addplot+ [error bars/.cd, y dir=both,y explicit,]
          table [x index=0, y index=3, y error index=4]{code/SIMD/pids_2D_flux_shima_etal_turbo_relative.dat};

        \addplot+ [error bars/.cd, y dir=both,y explicit,]
          table [x index=0, y index=5, y error index=6]{code/SIMD/pids_2D_flux_shima_etal_turbo_relative.dat};
      \end{axis}
    \end{tikzpicture}%
    \caption{2D, flux of Shima \etal \eqref{eq:flux_shima_etal}.}
  \end{subfigure}%
  \hspace*{\fill}
  \begin{subfigure}{0.49\linewidth}
    \begin{tikzpicture}[
      font=\footnotesize
      ]
      \begin{axis}[
          xlabel={Polynomial degree},
          ylabel={Relative \PID with SIMD},
          width=\textwidth,
          height=0.66\textwidth,
          ymin=0.0,
          grid=major,
        ]
        \addplot+ [error bars/.cd, y dir=both,y explicit,]
          table [x index=0, y index=1, y error index=2]{code/SIMD/pids_3D_flux_shima_etal_turbo_relative.dat};

        \addplot+ [error bars/.cd, y dir=both,y explicit,]
          table [x index=0, y index=3, y error index=4]{code/SIMD/pids_3D_flux_shima_etal_turbo_relative.dat};

        \addplot+ [error bars/.cd, y dir=both,y explicit,]
          table [x index=0, y index=5, y error index=6]{code/SIMD/pids_3D_flux_shima_etal_turbo_relative.dat};
      \end{axis}
    \end{tikzpicture}%
    \caption{3D, flux of Shima \etal \eqref{eq:flux_shima_etal}.}
  \end{subfigure}%
  \\
  \begin{subfigure}{0.49\linewidth}
    \begin{tikzpicture}[
      font=\footnotesize
      ]
      \begin{axis}[
          xlabel={Polynomial degree},
          ylabel={Relative \PID with SIMD},
          width=\textwidth,
          height=0.66\textwidth,
          ymin=0.0,
          grid=major,
        ]
        \addplot+ [error bars/.cd, y dir=both,y explicit,]
          table [x index=0, y index=1, y error index=2]{code/SIMD/pids_2D_flux_ranocha_turbo_relative.dat};

        \addplot+ [error bars/.cd, y dir=both,y explicit,]
          table [x index=0, y index=3, y error index=4]{code/SIMD/pids_2D_flux_ranocha_turbo_relative.dat};

        \addplot+ [error bars/.cd, y dir=both,y explicit,]
          table [x index=0, y index=5, y error index=6]{code/SIMD/pids_2D_flux_ranocha_turbo_relative.dat};
      \end{axis}
    \end{tikzpicture}%
    \caption{2D, flux of Ranocha \eqref{eq:flux_ranocha}.}
  \end{subfigure}%
  \hspace*{\fill}
  \begin{subfigure}{0.49\linewidth}
    \begin{tikzpicture}[
      font=\footnotesize
      ]
      \begin{axis}[
          xlabel={Polynomial degree},
          ylabel={Relative \PID with SIMD},
          width=\textwidth,
          height=0.66\textwidth,
          legend to name=legend:SIMD-relative-Euler,
          legend columns=-1,
          ymin=0.0,
          grid=major,
        ]
        \addplot+ [error bars/.cd, y dir=both,y explicit,]
          table [x index=0, y index=1, y error index=2]{code/SIMD/pids_3D_flux_ranocha_turbo_relative.dat};
          \addlegendentry{\trixi, \codeinline{TreeMesh}}

        \addplot+ [error bars/.cd, y dir=both,y explicit,]
          table [x index=0, y index=3, y error index=4]{code/SIMD/pids_3D_flux_ranocha_turbo_relative.dat};
          \addlegendentry{\trixi, \codeinline{StructuredMesh}}

        \addplot+ [error bars/.cd, y dir=both,y explicit,]
          table [x index=0, y index=5, y error index=6]{code/SIMD/pids_3D_flux_ranocha_turbo_relative.dat};
          \addlegendentry{\trixi, \codeinline{P4estMesh}}
      \end{axis}
    \end{tikzpicture}%
    \caption{3D, flux of Ranocha \eqref{eq:flux_ranocha}.}
  \end{subfigure}%
  \caption{Relative runtime improvements obtained by SIMD optimizations of the
           volume terms for different mesh types and LGL-DGSEM discretizations for the compressible Euler equations.}
  \label{fig:SIMD-relative-Euler}
\end{figure}

Figure~\ref{fig:SIMD-relative-Euler} shows the relative \PID improvements
obtained by SIMD optimizations of the flux differencing volume terms, i.e.,
the ratio of the \PID from Figures~\ref{fig:SIMD-PID-Euler} and
\ref{fig:Cartesian-vs-curved-PID-Euler}.
The trends described above are even more visible here. In particular, the speedup
obtained by all optimizations (precomputing terms and using SIMD optimizations)
improves for low polynomial degrees. In 2D, the speedup shows a local optimum at
$\polydeg = 7$.

Note that the \PID measures also the time needed for other parts of the right-hand
side computations. In particular, the cost of surface terms increases relatively
as the volume terms become cheaper.
Some SIMD optimizations could also be applied to surface terms to further increase
the total speedup. However, we restrict our attention to the flux differencing
volume terms in this article.

In total, precomputing variables and using SIMD optimizations improves the \PID
in 2D between
\SI{20}{\percent} (low polynomial degree, cheap volume flux) and
\SI{3}{\times} ($\polydeg = 7$, EC volume flux).
In 3D, the speedup obtained by these optimizations is between
\SI{30}{\percent} (low polynomial degree, cheap volume flux) and
\SI{3}{\times} (high polynomial degree, EC flux).

\section{Summary and conclusions}
\label{sec:summary}

We discussed techniques for the efficient implementation of flux differencing
schemes, focusing on the compressible Euler equations and discontinuous Galerkin
methods. Starting with a high-level description of the algorithms, we presented
general modifications of the equations typically presented in research articles
as first step towards an efficient implementation. All of these techniques are
freely available in our open source codes \trixi and \fluxo as well as our
reproducibility repository \cite{ranocha2021efficientRepro}.

We concentrated on the serial performance of flux differencing for the
compressible Euler equations in 2D and 3D. Most of the techniques presented in
this article are agnostic to the code base and programming language, as
demonstrated by results obtained with Julia and Fortran.
Extensions to non-conservative terms as well as MPI parallelization will be
discussed elsewhere, since these questions are largely orthogonal to the issues
discussed here. In the MPI parallel case, one
would usually expect a sublinear scaling up to a full single node due to memory
bandwidth and cache competition followed by a (close to) ideal scaling when adding
more nodes, as shown in \cite{rogowski2022performance}.

From these general performance optimizations, we compared flux differencing
to a simple version of overintegration. In general, flux differencing is quite
competitive in terms of runtime performance. In addition, it comes with less
strict explicit time step restrictions and is robust for many setups where overintegration
fails \cite{winters2018comparative}. For practically relevant parameters for
computational fluid dynamics (3D, polynomials of degree $p = 3$), flux differencing
is even faster than overintegration with a single additional node per coordinate
direction.

We also discussed more invasive optimizations including memory layout adaptation for SIMD
techniques. While these are rather code-specific, they can provide great
benefits on modern hardware. Using a compromise of legacy code layout and SIMD
optimizations, we could achieve
the following performance indices \PID
(time per right-hand side evaluation and degree of freedom)
for flux differencing discretizations of the compressible Euler equation in 3D
with polynomials of degree $\polydeg = 3$ in Trixi.jl
on an Intel\textregistered\ Core\texttrademark\ i7-8700K (CPU from 2017 with AVX2):
\SI{2.6e-8}{\s} on a Cartesian mesh and
\SI{4.3e-8}{\s} on an unstructured, curved mesh
for standard split form discretizations using the flux of Shima \etal \cite{shima2021preventing};
\SI{4.2e-8}{\s} on a Cartesian mesh and
\SI{6.5e-8}{\s} on an unstructured, curved mesh
for entropy-based discretizations using the flux of Ranocha \cite{ranocha2018comparison,ranocha2018thesis,ranocha2020entropy}.

%% file: acknowledgments.tex
Funded by the Deutsche Forschungsgemeinschaft (DFG, German Research Foundation)
under Germany's Excellence Strategy EXC 2044-390685587, Mathematics M\"{u}nster:
Dynamics-Geometry-Structure, and through the DFG research unit ``SNuBIC'' (FOR 5409; project number 463312734).

Hendrik Ranocha was supported by the Daimler und Benz Stiftung (Daimler and Benz
foundation, project number 32-10/22).

This work has received funding from the European Research Council through the
ERC Starting Grant ``An Exascale aware and Un-crashable Space-Time-Adaptive
Discontinuous Spectral Element Solver for Non-Linear Conservation Laws'' (Extreme),
ERC grant agreement no. 714487 (Gregor J. Gassner, Michael Schlottke-Lakemper, and Andr\'{e}s Rueda-Ram\'{i}rez).

Andrew R. Winters was supported through Vetenskapsr{\aa}det, Sweden grant
agreement 2020-03642 VR.

Jesse Chan was supported through the United States National Science Foundation
under awards DMS-1719818 and DMS-1943186.

The authors gratefully acknowledge the computing time provided on the supercomputer NEC Vulcan by
the High-Performance Computing Center Stuttgart (HLRS) of the University of Stuttgart, Germany.